\documentclass[final,authoryear,5p,times,twocolumn]{elsarticle}

\usepackage{epsfig}

\usepackage{amssymb}

\usepackage{lineno}
\usepackage{rotating}
\usepackage{color,soul}

\usepackage[bookmarks = true, bookmarksnumbered = true, pdfpagemode =None, pdfstartview = FitH, pdfpagelayout = SinglePage, colorlinks = true, urlcolor = blue, citecolor = monbleu]{hyperref}

\newcommand{\pderiv}[2]{\frac{\partial #1}{\partial #2}}

\newcommand{\beq}{\bigskip\begin{equation}}
\newcommand{\eeq}{\bigskip\end{equation}}

\journal{Icarus}

\begin{document}

\begin{frontmatter}



\title{A dispersive wave pattern on Jupiter's fastest retrograde jet at 20$^\circ$S}


\author[jr]{J.H. Rogers}
\ead{jrogers11@btinternet.com}

\author[le]{L.N. Fletcher}

\author[ga]{G. Adamoli}
\author[mj]{M. Jacquesson}
\author[mv]{M. Vedovato}
\author[go]{G.S. Orton}
%

\address[jr]{British Astronomical Association, Burlington House, Piccadilly, London  W1J 0DU, UK}
\address[le]{Department of Physics \& Astronomy, University of Leicester, University Road, Leicester, LE1 7RH, UK}
\address[ga]{JUPOS Team, c/o British Astronomical Association}
\address[mj]{JUPOS Team, c/o British Astronomical Association}
\address[mv]{JUPOS Team, Unione Astrofili Italiani}
\address[go]{Jet Propulsion Laboratory, California Institute of Technology, Pasadena, CA 91109, USA.}


\begin{abstract}
A compact wave pattern has been identified on Jupiter's fastest retrograding jet at $20^\circ$S (the SEBs) on the southern edge of the South Equatorial Belt.  The wave has been identified in both reflected sunlight from amateur observations between 2010 and 2015, thermal infrared imaging from the Very Large Telescope and near infrared imaging from the Infrared Telescope Facility.  The wave pattern is present when the SEB is relatively quiescent  and lacking large-scale disturbances, and is particularly notable when the belt has undergone a fade (whitening).  It is generally not present when the SEB exhibits its usual large-scale convective activity (`rifts'). Tracking of the wave pattern and associated white ovals on its southern edge over several epochs have permitted a measure of the dispersion relationship, showing a strong correlation between the phase speed (-43.2 to -21.2 m/s) and the longitudinal wavelength, which varied from $4.4-10.0^\circ$ longitude over the course of the observations. Infrared imaging sensing low pressures in the upper troposphere suggest that the wave is confined to near the cloud tops.  The wave is moving westward at a phase speed slower (i.e., less negative) than the peak retrograde wind speed (-62 m/s), and is therefore moving east with respect to the SEBs jet peak.  Unlike the retrograde NEBn jet near $17^\circ$N, which is a location of strong vertical wind shear that sometimes hosts Rossby wave activity, the SEBs jet remains retrograde throughout the upper troposphere, suggesting the SEBs pattern cannot be interpreted as a classical Rossby wave.  2D windspeeds and thermal gradients measured by Cassini in 2000 are used to estimate the quasi-geostrophic potential vorticity gradient as a means of understanding the origin of the a wave.   We find that the vorticity gradient is dominated by the baroclinic term and becomes negative (changes sign) in a region near the cloud-top level (400-700 mbar) associated with the SEBs.  Such a sign reversal is a necessary (but not sufficient) condition for the growth of baroclinic instabilities, which is a potential source of the meandering wave pattern.
\end{abstract}


\end{frontmatter}


\section{Introduction}
\label{intro}

Jupiter's jets have generally been regarded as straight and fixed in latitude \citep{86limaye, 95rogers, 05vasavada}, especially the main prograde (eastward) jets and the fastest of the retrograde (westward) jets.  However, substructure such as vortices, eddies and wave-like features, along with complex temporal variability, are increasingly being discovered within them \citep[e.g.,][]{08sanchez, 08rogers_jup07, 11asaydavis, 12simon}.  Waves are diagnostic of the background medium in which they propagate, so if specific types of wave patterns could be identified, measurements of their parameters could constrain the three-dimensional profiles of jets.  A number of regular wave patterns in the visible cloud deck have been identified at several latitudes.  Those of shortest wavelength (approximately $\sim300$ km, i.e. mesoscale waves), were observed by Voyager, Cassini and New Horizons as packets of bright bands lying roughly north-south across the equatorial zone. They have most commonly been interpreted as gravity waves, but may alternatively be gravity-inertia waves or Kelvin waves \citep{79hunt, 86flasar, 09arregi, 15simon_grav}.  Long wave-trains of similar appearance but longer wavelength (700-1200 km; $0.6-1.0^\circ$ longitude) have also been recorded on the North Equatorial Belt (NEB) at $14-17^\circ$N, on the cyclonic flank of the retrograde jet, in recent HST images and in some Voyager images; they have been interpreted as baroclinic waves \citep{15simon}. (All latitudes quoted in this paper are planetographic, and all longitudes refer to System III west).

A variety of waves have been observed on the prograde jet at $7^\circ$S (the northern edge of the South Equatorial Belt, the SEBn).  These include the chevron-shaped dark features, usually visible around most or all of the circumference with a spacing of $4-8^\circ$ longitude, for which an inertia-gravity wave model has been favoured \citep{90allison, 12simon}.  In addition there is sometimes a large, long-lived solitary disturbance, which has been modelled as a Rossby wave \citep{90allison, 12simon}.  Animations of Cassini maps revealed a remarkable meridional oscillation of the faster-moving chevrons east of this large feature, tracing out a pattern of wavelength $20^\circ$, which was also interpreted as a Rossby wave \citep{12simon}. Evidently, multiple wave modes influence the motions of features on this prograde jet.
	
Two types of large-scale waves have been observed on the north and south sides of the NEB. Despite the many differences between these two wave patterns, both have been interpreted as Rossby waves.  The most striking of these waves, with wavelengths of $\approx~20-40^\circ$ longitude, are thought to be responsible for the visibly dark formations also known as 5-$\mu$m `hotspots,' regions of anomalously bright infrared emission that dominate the prograde jet at $7^\circ$N.  These are likely to be examples of an Equatorial Rossby Wave \citep[e.g.,][]{90allison, 98ortiz, 05friedson, 06arregi}, and may have been the source of the unexpected compositional results from the Galileo probe \citep{98niemann,98orton}.  

Rossby wave activity is also thought to have been detected on the retrograde jet at $\sim17^\circ$N (the northern edge of the North Equatorial Belt, NEBn), although the arrangement was very different from that on the equatorial jets (see Section \ref{discuss} for discussion).  Unlike the waves on the equatorial jets, which have characteristic phase speeds and create large-scale disturbances of the main cloud deck, the NEBn wave pattern was almost stationary in System III and featured a diffuse variation of haze thickness and temperature, only detectable above the main cloud deck.  Such variations were originally identified as slowly-moving thermal waves above the cloud-tops near 250 mbar, which \citet{94orton} and \citet{97deming} attributed to Rossby waves that could arise as low-amplitude meridional excursions of the NEBn jet at cloud-top level and propagate upwards as thermal waves.  Such waves were most regular and conspicuous in 2000 -- fortuitously, during the Cassini flyby -- when they were observed simultaneously in ultraviolet images, strong-methane-band images and thermal maps derived from Cassini \citep{03porco, 04flasar_jup, 06li}, and in ground-based methane-band images \citep{04rogers}. The Rossby wave interpretation was developed by \citep{06li}, suggesting a wave confined to the upper troposphere (200-400 mbar) with a strong coupling of the opacity and the thermal field through vertical motions. However, concurrent ground-based observations showed that this wave pattern with a wavelength of $20-25^\circ$ longitude was coupled to, and probably controlled by, large-scale visible circulations flanking the NEBn jet \citep{04rogers}.

In contrast, here were report for the first time wave activity on the retrograde jet at the southern edge of the Southern Equatorial Belt (SEBs, $20^\circ$S).  No large-scale wave activity has been previously recorded on the SEBs, but as ground-based imaging achieves higher and higher resolution, smaller scales comparable to the jet width ($\approx1-2^\circ$) are becoming accessible.  

The SEBs and the NEBn jets are different from one another: the SEBs jet is by far the fastest retrograde jet\footnote{In this paper, `fast' and `slow' refer to speeds relative to System III, so in the retrograding SEBs jet `fast' refers to positive drift in longitude (westward) or negative wind speed ($u$).  All latitudes quoted in this paper are planetographic, and all longitudes refer to System III west.} on Jupiter, is influenced by the presence of the Great Red Spot (no similar large vortex is associated with the NEBn jet), and is the only one that carries mid-scale vortices (several thousand kilometres in diameter) that have been tracked historically \citep{95rogers}.  In these respects it appears comparable to the prograde jets that carry similar vortices.  From the Cassini zonal wind profile (ZWP) \citep{03porco}, which we generally adopt for reference, it has a peak speed of $u=-62.0$ m/s at $19.7^\circ$S.  

The SEBs jet usually carries some mid-scale anticyclonic vortices on the south edge of the peak, as seen in Voyager and Cassini images \citep{95rogers, 03porco}.  They are also visible in ground-based images over many years, identified as vortices by their oval shape, higher-albedo central spot, and latitude ($\approx21^\circ$S).  They are located $1^\circ$ south of the nominal jet peak, and retrograde with almost the full speed of the jet, typically with $u$ ranging from $\sim$-53 m/s to -64 m/s \citep{08rogers_jup07, 15rogers_jup05}.  So, assuming the jet peak position quoted above, the measurements support the paradigm established for some prograde jets \citep{05garcia} that these spots are vortices which `roll' along the anticyclonic side of the jet peak, with the smallest, fastest ones being close to the peak, but the majority being slightly slower by ~5-15 m/s.

Here we report the discovery of a regular wave pattern in the peak of the SEBs retrograde jet, observed in several recent years, with a phase speed that is much less than the retrograde wind speed.  The observations from visible and thermal infrared wavelengths are described in Section \ref{obs}.  Section \ref{results} shows that the phase speed is closely correlated with the longitudinal wavelength, and Section \ref{discuss} interprets these results in terms of jovian wave activity and compares them to other wave systems reported on the giant planets.

\section{Observations of the SEBs Wave}
\label{obs}

\subsection{Visible Light Observations}
\label{vis}
\begin{figure*}
\begin{centering}
\centerline{\includegraphics[angle=0,scale=1.0]{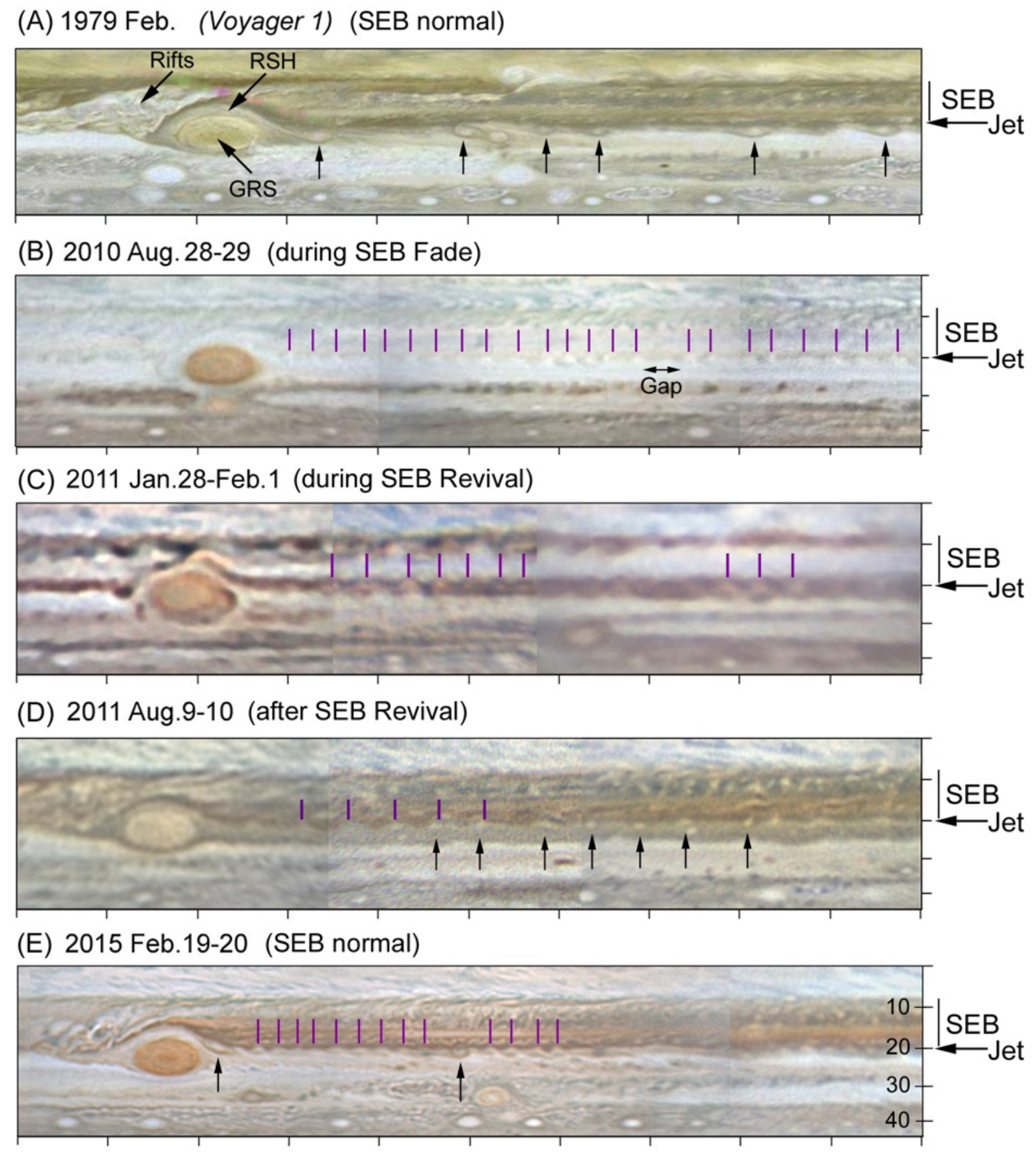}}
\caption{Cylindrical map projections showing the SEBs wave pattern in each of the four epochs, and a Voyager map for comparison. Each map shows latitudes $0-45^\circ$S, covering $200^\circ$ longitude, with longitude gradations at $20^\circ$ intervals. All maps were made by M. Vedovato (JUPOS project) apart from the Voyager map.  Rows of purple marks above the SEBs indicate the slow-moving wave-train.  Black arrows below the SEBs indicate features moving at or near full jet speed. Labels at right indicate the position of the SEB, the SEBs jet (arrows), and planetographic latitudes (shown in last panel). Labels in the first panel indicate the Great Red Spot (GRS), Red Spot Hollow (RSH), and ÔriftsÕ in the SEB.  (A)  1979 Feb.  Voyager 1.  (NASA image PIA00011.)  There are numerous vortices retrograding close to full jet speed, but no visible waves.  (B)  2010 Aug.28-29, during the SEB Fade. Images by A. Wesley, T. Ikemura, and K. Yunoki. Double-headed arrows show a gap in the wave-train, moving with full jet speed. (C)  2011 Jan.28-Feb.1, during the SEB Revival.  Images by G. Walker and G. Jolly. Resolution was lower at this time, towards the end of the apparition.  The wave-train is seen as a chain of light spots within the revived dark southern component of the SEB. (By this stage of the Revival, all the dark spots retrograding on this belt had moved to its south edge and were moving with approximately the same speed as the wave-train.) (D)  2011 Aug.9-10, after the SEB Revival.  Images by D. Peach, T. Hasebe, and K. Tokujiro (by courtesy of the ALPO-Japan).  Arrows mark small white spots along the SEBs, spaced $12-16^\circ$ apart, moving through the wave-train with full jet speed.  (E)  2015 Feb.19-20, with SEB normal.  Images by T. Olivetti and T. Horiuchi.  Arrows mark two vortex rings with full jet speed, one of them entering the Red Spot Hollow. }
\label{cmaps}
\end{centering}
\end{figure*}

\begin{figure*}
\begin{centering}
\centerline{\includegraphics[angle=0,scale=1.0]{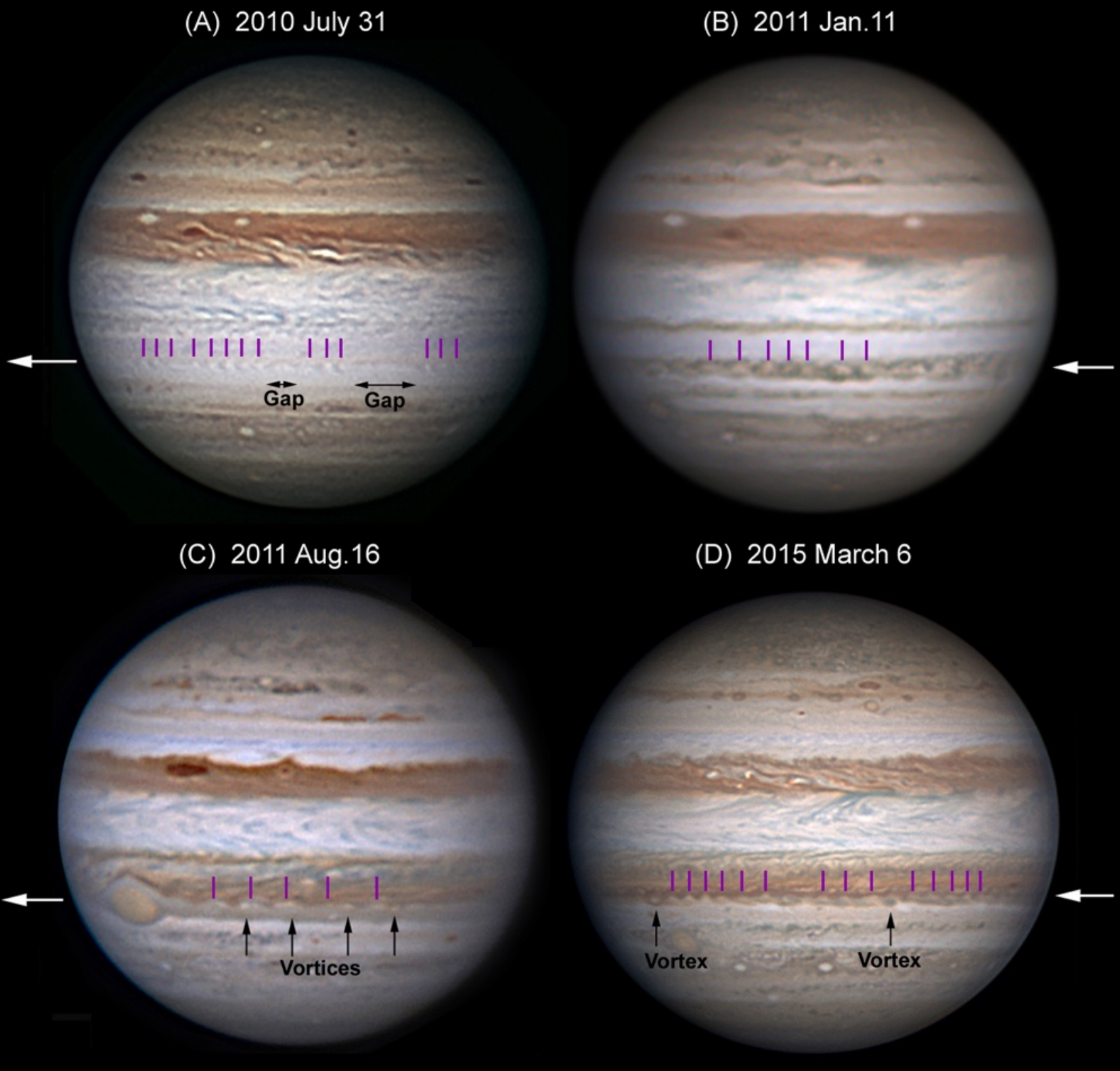}}
\caption{Images showing the SEBs wave pattern in each of the four epochs. White arrows indicate the latitude of the SEBs (pointing west, in the direction of motion). Rows of purple marks above the SEBs indicate the slow-moving wave-train.  Black arrows below the SEBs indicate features moving at or near full jet speed.  The images were acquired as follows: (A)  2010 July 31, 09:08 UT, during the SEB Fade.  Image by Gary Walker.  Double-headed arrows mark gaps in the wave-train, moving with full jet speed. (B)  2011 Jan.11, 23:53 UT, during the SEB Revival.  Image by Donald C. Parker.  The wave-train is seen as a chain of white ovals within the revived dark southern component of the SEB. (C)  2011 Aug.16, 01:42 UT, after the SEB Revival.  Image by John Rozakis. Arrows mark small white spots along the SEBs, spaced $12-16^\circ$ apart, moving through the wave-train with full jet speed. (D)  2015 March 6, 15:44 UT, with SEB normal.  Image by Tiziano Olivetti. Arrows mark two spots with full jet speed: a vortex ring (left) and a small dark oval (right).}
\label{images}
\end{centering}
\end{figure*}

\begin{figure*}
\begin{centering}
\centerline{\includegraphics[angle=0,scale=0.7]{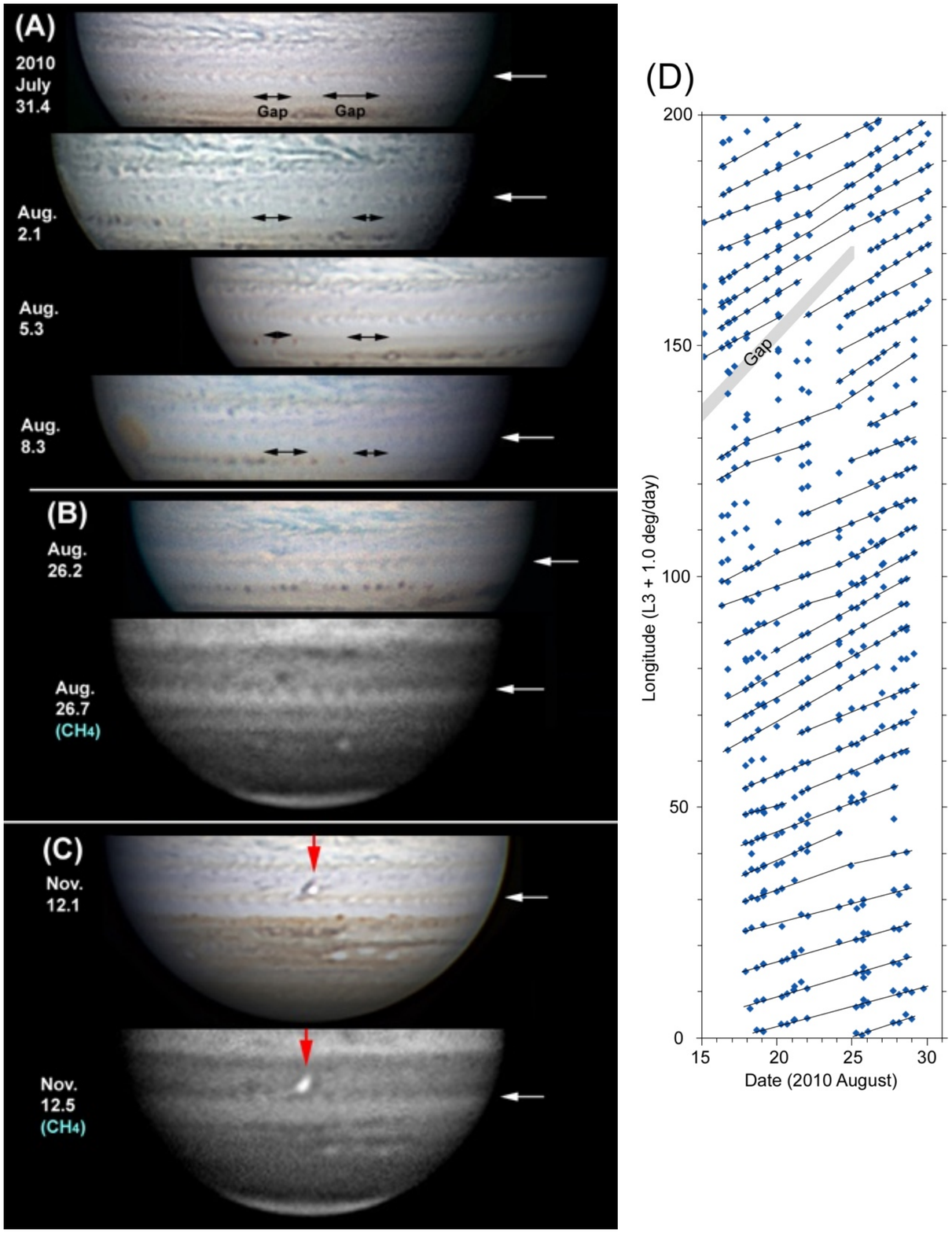}}
\caption{Sequence of images showing the SEBs wave pattern during the SEB fade in 2010. White arrows mark the latitude of the SEBs jet. (A)  Sequence of images in 2010 August, showing the wave pattern with two of the rapidly-moving short gaps (indicated as black horizontal arrows) in which the pattern is indistinct.  (B)  Image in visible light paired with one in the 0.89 micron methane absorption band taken one rotation later.  In the methane band, which is sensitive to aerosols above the main cloud deck, the SEB is still dark and the SEBs wave pattern has high contrast.  (C)  Images on 2010 Nov.12, showing the SEBs wave pattern in visible light and methane band (0.89 microns), for comparison with the thermal infrared image in Fig. \ref{visir}. The initial outbreak of the SEB Revival is also visible (marked with red arrow), but has not yet affected the SEBs more than a few degrees away.  Images were taken by T. Akutsu (890 nm), B. Combs, W. Jaeschke, D.C. Parker, G. Tarsoudis, and G. Walker. (D) Excerpt from the JUPOS chart for SEBs features in 2010 August, plotted in a longitude system moving at $+1.0$ deg/day relative to System III.  Only dark features are plotted; spaces between them are the bright spots.  The grey band indicates one of the gaps moving with full jet speed as identified on the complete chart over a longer time-span; tracks of the wave features are disrupted or missing within it.}
\label{fade}
\end{centering}
\end{figure*}

Images acquired by amateur observers over the past several years (Fig. \ref{cmaps}-\ref{images}) have revealed a wave-like pattern on Jupiter's SEBs, in the latitude of the jet peak but with much slower retrograde speed (i.e., less negative than the zonal windspeed).  The wave marks the boundary between the SEB and South Tropical Zone (STrZ), and sometimes features white ovals to the south side of the SEBs jet, nestled between the wave crests.  These waves coexisted with other features in overlapping latitudes that moved with the usual peak speed of the jet (Fig. \ref{cmaps}).  This phenomenon was first evident during the cycle of SEB fading and revival in 2010-11, which involved a SEB Fade \citep[quiescence and whitening of the normally dark belt,][]{11fletcher_fade,12perezhoyos} followed by vigorous convective and turbulent disturbances that comprise the SEB Revival \citep{13giles,11rogers_epsc}.  Occurrences of the wave at different phases of the SEB lifecycle will be described below.  Four different epochs are displayed as global images in Fig. \ref{images} and as cylindrical maps (involving images from multiple observers) in Fig. \ref{cmaps}.  

The observations consist of ground-based images obtained by numerous amateur astronomers with telescopes of $20-41$ cm aperture, using the webcam image stacking technique \citep{14mousis_proam, 15kardasis}. These have been analysed as part of the routine surveying of the planet by the JUPOS team and the British Astronomical Association\footnote{http://www.britastro.org/jupiter/section\_reports.htm}.  Positions of `spots' were measured by the JUPOS project \citep{15kardasis}.

Both colour images and red-light images are used, with processing done by each observer to enhance small-scale detail and contrast.  The spatial resolution is typically $0.5-1.0^\circ$ on the planet.  Measurements of `spots' are done on-screen using the WinJUPOS program (created by G. Hahn), which is fully described and available at \mbox{http://jupos.org}. In WinJUPOS, the measurer carefully fits each image to an elliptical outline frame, after brightening to reveal the limb clearly, and the alignment may be checked by reference to positions of satellites and their shadows if present, and to positions of well-known features on Jupiter's disc. The measurer then positions the cursor over every distinct feature (`spot') in turn and its longitude and latitude are automatically displayed, and can be recorded into the database.  As an example, for the apparition of 2010-11, many hundreds of images were measured and the database contained 105,118 measurements over the whole planet. 

Measurements of an individual spot typically have a standard deviation of about $0.4^\circ$ in latitude, which can therefore be taken as the uncertainty of measurement in latitude or longitude.  To establish latitudes, at least 5 points (and usually many more, from images by multiple observers on multiple dates) are averaged for each spot or group of spots; therefore the standard error of latitudes is less than $0.2^\circ$.

Data are then selected by latitude range and plotted in WinJUPOS as longitude-vs-time charts (e.g., Fig. \ref{fade}D), on which individual spots are evident as linear arrays of points.  In the case of the closely-spaced wave-like features described in this paper, care was taken to use sufficient images to avoid aliasing. Zonal drift rates are then measured in degrees per 30 days and converted to zonal windspeed ($u$).  The drift rates are derived from the charts by one of two methods in WinJUPOS, which give equivalent results: by least-squares fit; or by fitting a line by eye (drift rate being displayed on screen).  The individual `spots' which comprise the wave-trains described herein were usually tracked for 10-30 days each (in 2010 and 2011 August), or 5-20 days each (in 2011 January and in 2015), with $\sim1$ measurement every 2 days in 2011 August and $\sim1$ measurement per day in the other three epochs. Given the typical uncertainty of $\pm0.4^\circ$ for individual measurements, if a spot is tracked with constant speed for 10 days, the uncertainty in drift rate is $\pm(2\times0.4^\circ)$ per 10 days, or $\pm1.1$ m/s.  For groups of spots (the wave-trains in Table \ref{waveparams}), the quoted uncertainties are either the standard deviation of the individual tracks within the wave-train, or the range of acceptable fits to the wave-train as a whole.\footnote{The observations and measurements for each epoch are fully described in on-line reports, as follows: 2010 May-Nov:  2010/11 reports 8 and 22 \mbox{(http://www.britastro.org/jupiter/2010reports.htm)}; 2011 Jan.: 2010/11 reports 22 and 24 \mbox{(http://www.britastro.org/jupiter/2010reports.htm)}; 2011 July-August:  2011/12 final report, \mbox{(http://www.britastro.org/jupiter/2011report09.htm)}; and 2014 Oct.-2015 Mar.: 2014/15 report \mbox{http://www.britastro.org/jupiter/2014\_15report10.htm}.}

Slow-moving wave-like patterns were observed in three very different states of the SEB in 2010-2011 (Fig. \ref{cmaps} and Fig. \ref{images}): (1) during the SEB Fade in summer 2010 (Fig. \ref{cmaps}B and \ref{images}A), when the SEB was quiet and whitened, but gaps in the wave-train (Figs. \ref{cmaps}, \ref{images}, \ref{fade}) moved at full jet speed; (2) during the SEB Revival in late 2010 and early 2011 (Fig. \ref{cmaps}C and \ref{images}B), when the wave-train reappeared even as dark spots from the vigorous Revival moved with full jet speed; (3) when the SEB was returning to normal in late 2011 (Fig. \ref{cmaps}D and \ref{images}C), and the wave-train co-existed with small bright spots at full jet speed.  Fourthly, the last situation has been observed again in 2014-15, with the SEB in its normal state (Fig. \ref{cmaps}E and \ref{images}D).  Fig. \ref{cmaps} also compares the 2010-2015 images to those acquired by Voyager 1 in February 1979, where there is no evidence for a similar wave train on the SEBs. These four epochs will be described in turn below, and Table \ref{sequence} summarises the status of the SEB during these times.

\begin{table*}[htdp]
\caption{Time-line of relevant changes in the SEB when the SEBs wave pattern was detected.  Epochs refer to the itemised list in the main text.}
\centering
\begin{tabular}{|l|l|}
\hline
Time	& Event \\
\hline
2009 Jul-Aug.	& SEB Fade beginning; convective activity near the GRS had ceased. \\
2010 May-Nov &  Observation of SEBs wave pattern (epoch 1) \\
2010 Nov.	& SEB Revival: initial outbreak on Nov. 9, then spreading east \\
		& in mid-SEB and west on SEBs jet, wave pattern still present. \\
2011 Jan.	& SEB Revival (continued): disturbance at all longitudes;  \\
		& wave pattern reappears in sectors of revived belt (epoch 2). \\
2011 April	&  Solar conjunction. \\
2011 May	& SEB fully revived but normal convective disturbances near GRS \\
		& had not resumed.  Wave pattern observable in July-Aug. (epoch 3). \\
2011 Sep.	& Rifting resumes to the west of the GRS on Sep 21. \\
2012-2013 &	Normal SEB activity, but no comparable wave pattern detectable. \\
2013-2014 &	Normal SEB activity, but no wave pattern detectable. \\
		
2014-2015 &	Normal SEB activity: wave pattern reappears (epoch 4) \\
\hline
\end{tabular}
\label{sequence}
\end{table*}%

\begin{enumerate}

\item \textbf{SEB Fade:}  The wave pattern was first observed in 2010, from May onwards, with the SEB fully faded \citep[appearing almost as bright as the adjacent zone,][]{11fletcher_fade,12perezhoyos}.  It appeared as a chain of small white ovals separated by pale grey patches, with a spacing of $5-7^\circ$ longitude (Fig. \ref{cmaps}B and Fig. \ref{images}A). The chain extended all around the planet, emerging within a few degrees west of the Red Spot Hollow (RSH), and continuing until the spots slid into the RSH from the east side.  The latitude of these white spots was initially $20.9^\circ$S, which was close to the peak of the mean zonal wind speed as indicated by the Cassini ZWP.  But the white spots' latitude reduced over several months until in one sector it was as low as $20.2^\circ$S, clearly different from the mean zonal wind speed.  It is possible that this was due to changing size of the spots, which were only one aspect of a broad wave-train that was centred and fixed along their north edge at $19.5-20.0^\circ$S, as in subsequent instances.  This was indicated by methane-band and thermal-IR (8.6 micron) images (see below, Section \ref{thermalir}), in which the SEB was still clearly visible and the wave-train marked its southern edge (Fig. \ref{fade}).  The speed of the wave crests (relative to System III longitude) varied from $\sim-26$ to $\sim-40$ m/s in large longitude blocks over several months.  The pattern was interrupted by several short gaps, which moved with peak jet speed: $u=-65\pm3$ m/s.  The SEB Revival began on 2010 November 9, but the wave pattern remained visible for several months thereafter, at all longitudes where the SEBs was not yet affected. 

\item \textbf{SEB Revival:}  During the SEB Revival in late 2010, the wave pattern persisted even as very dark spots retrograded along it at full jet speed.  The complex, rapidly changing features of the Revival have been described in detail \citep[see reports listed in footnote 2 and][]{11rogers_epsc} and will be summarised in a separate paper (Fletcher, Rogers \& colleagues, in preparation).  Eventually,  complex dark spots reconstituted the dark SEBs in the same latitude. But despite this complex activity in January 2011, the wave pattern reappeared inside some sectors of the SEBs as a chain of white ovals spaced $\approx8^\circ$ apart (Fig. \ref{cmaps}C and Fig. \ref{images}B).  These chains had a speed of $-27$ m/s, retrograding at a speed smaller than the zonal jet itself.

\item \textbf{Normal SEB 2011:}  The wave pattern reappeared after solar conjunction, in 2011 July-August, when the SEB was largely normal, except that the usual convective disturbances adjacent to the GRS had not yet resumed.  In a sector $\sim15-75^\circ$ east of the GRS, two series of small white spots coexisted on the SEBs with very different speeds (Fig. \ref{cmaps}D and Fig. \ref{images}C. One was a slowly retrograding set of white spots interspersed with dark southward bulges (spaced $10^\circ$ longitude apart), which from their sinusoidal appearance and slow motion are interpreted as a wave-train; the other was a rapidly retrograding series of discrete white spots (spaced $12-16^\circ$ longitude apart) which from their appearance as unconnected tiny ovals, and their rapid speed, are interpreted as small vortices. Interactions could be seen as vortices squeezed along the wave pattern.  The wave pattern was not seen after August 2011.


\item \textbf{Normal SEB 2014-15:}  In 2014-15, a similar pattern reappeared, after two apparitions (2012-13 and 2013-14) when regular wave-trains were only present occasionally if at all.  However, the modestly-retrograding wave-train had a spacing of $\sim5^\circ$ rather than the larger $10^\circ$ spacing of 2011, and a different speed.  In 2015 the wave pattern was visible over $\sim180^\circ$ of longitude east of the GRS (Fig. \ref{cmaps}E and Fig. \ref{images}D), but not continuously; it mainly comprised limited wave-trains east of the more typical jetstream vortices, which were moving with full jet speed ($u=-55$ to $u=-64$ m/s) slightly to the south of the jet peak.  So in 2015 the waves appeared to arise in the 'wake' of the vortices.  This relationship had not been observed in previous years but we had not previously observed such substantial vortices coexisting with a slower wave pattern.

\end{enumerate}

Having described the general appearance of the wave train and vortices in this section, we will use the measurements of the phase speed, meridional amplitude and wavelength from all four epochs to study the properties of the wave in Section \ref{results}, comparing it to the zonal wind profile.

\subsection{Thermal Infrared Observations}
\label{thermalir}

\begin{figure}[htdp]
\begin{centering}
\centerline{\includegraphics[angle=0,scale=0.9]{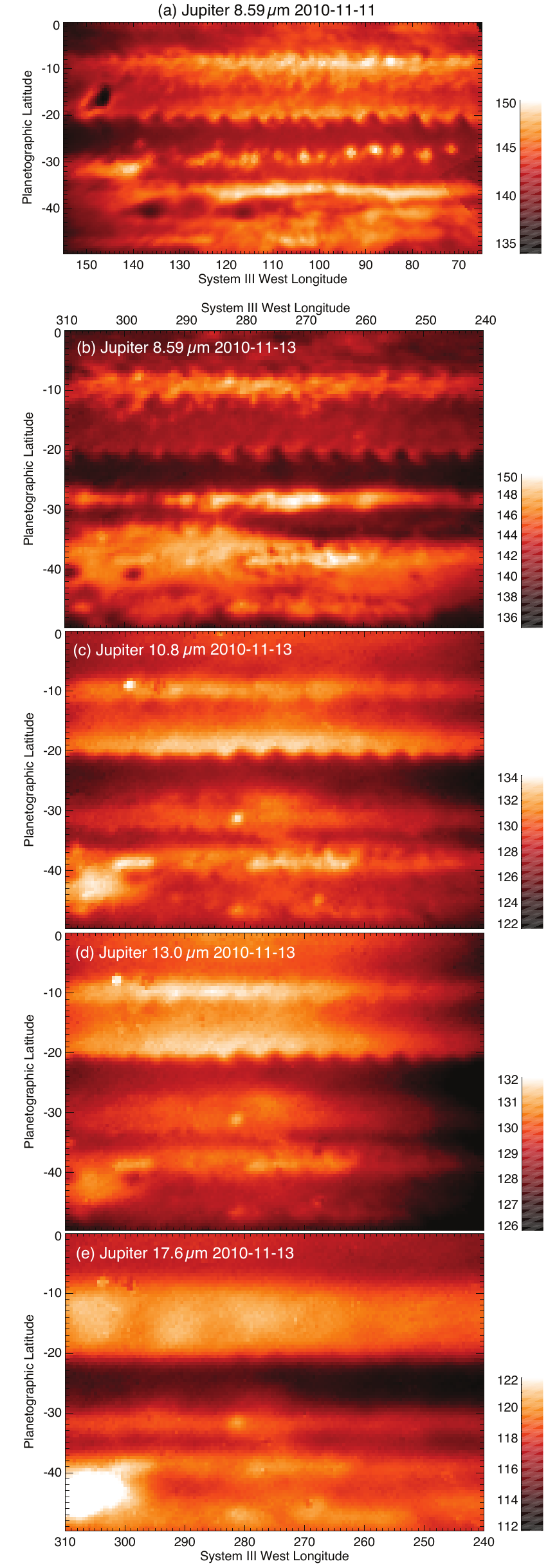}}
\caption{Thermal-infrared brightness temperature maps generated from VLT/VISIR images revealing the SEBs wave at $20^\circ$S.  Panel (a) shows a single 8.6-$\mu$m image obtained on Nov 11 2010 at 23:56UT, showing the revival plume at $150^\circ$W as a dark region of high aerosol opacity.  Panels b-e are from Nov 13 2010 for the opposite hemisphere (centred on $275^\circ$W), and show that the wave is clearly visible between 8-13 $\mu$m, but hard to see at 17.6 $\mu$m sensing the upper troposphere.  These images were acquired in program 286.C-5009(A). }
\label{visir}
\end{centering}
\end{figure}

Thermal emission at mid-infrared wavelengths (7-25 $\mu$m) is sensitive to the atmospheric temperatures and aerosol opacity in the 200-600 mbar region of Jupiter's troposphere, providing a second technique for investigating oscillations of the zonal jets. The fade and revival cycle of the SEB (Table \ref{sequence}) was studied extensively using photometric imaging by the VISIR instrument \citep{04lagage} on ESO's Very Large Telescope (VLT).  Images at 8.6 $\mu$m (Fig. \ref{visir}a-b), which are mostly sensitive to changes in tropospheric aerosol opacity near the 600-mbar level \citep[e.g.,][]{09fletcher_imaging}, provide excellent contrast between the SEB and the neighbouring Southern Tropical Zone (STrZ), and reveal the SEBs wave pattern as an undulation of the boundary between the two bands.  

By the time the SEB was fully faded in July 2010 (epoch 1), \citet{11fletcher_fade} identified undulations of the cloud opacity at 8.6 $\mu$m and $19-20^\circ$S with a $5-6^\circ$ longitude wavelength, which was the same wave-train that we have described above from amateur imaging.  No undulations of the SEBs had been noted in VISIR imaging acquired prior to October 2009, consistent with the timeline discussed above.  Here we report VISIR images at the time of the SEB revival on November 11th and 13th 2010 \citep{13giles}, when the SEBs undulations of epoch 1 were clearly still present before being disrupted by the chaotic rifting associated with the revival through 2011 (epoch 2).

Fig. \ref{visir}a-b shows two 8.6-$\mu$m images centred at $115^\circ$W and $275^\circ$W (taken 2.5 jovian rotations apart and covering two different hemispheres of Jupiter).  The undulation was also clearly visible at 10.7 $\mu$m (Fig. \ref{visir}c, sensitive to a combination of tropospheric temperatures and ammonia humidity) and 13.1 $\mu$m (Fig. \ref{visir}d, sensitive to 600-mbar temperatures).  The undulating SEBs is much harder to see at Q-band wavelengths (17.8, 18.6 and 19.5 $\mu$m, Fig. \ref{visir}e) sensing Jupiter's upper tropospheric temperatures (200-400 mbar), albeit at a lower diffraction-limited spatial resolution (diffraction limit of 0.56" at 17.6 $\mu$m, equivalent to 1690 km on Jupiter or $1.4^\circ$ at the equator).  This spatial resolution would have been sufficient to resolve the wave pattern (longitudinal spacings of $5-7^\circ$) if it were present in the 200-400 mbar region.  The wave must be present in the 400-600 mbar region where contribution functions of the 8.6-, 10.7- and 13.1-$\mu$m filters peak.  Like the NEBn wave observed at $17^\circ$N in 2000, the 7.9-$\mu$m images (sensitive to methane emission) obtained in 2010 did not show evidence of the SEBs wave at 1- to 5-mbar pressure levels.  We compare the NEBn and SEBs waves in more detail in Section \ref{discuss}.

\subsection{Near-infrared observations}

\begin{figure}[!htb]
\begin{centering}
\centerline{\includegraphics[angle=0,scale=0.4]{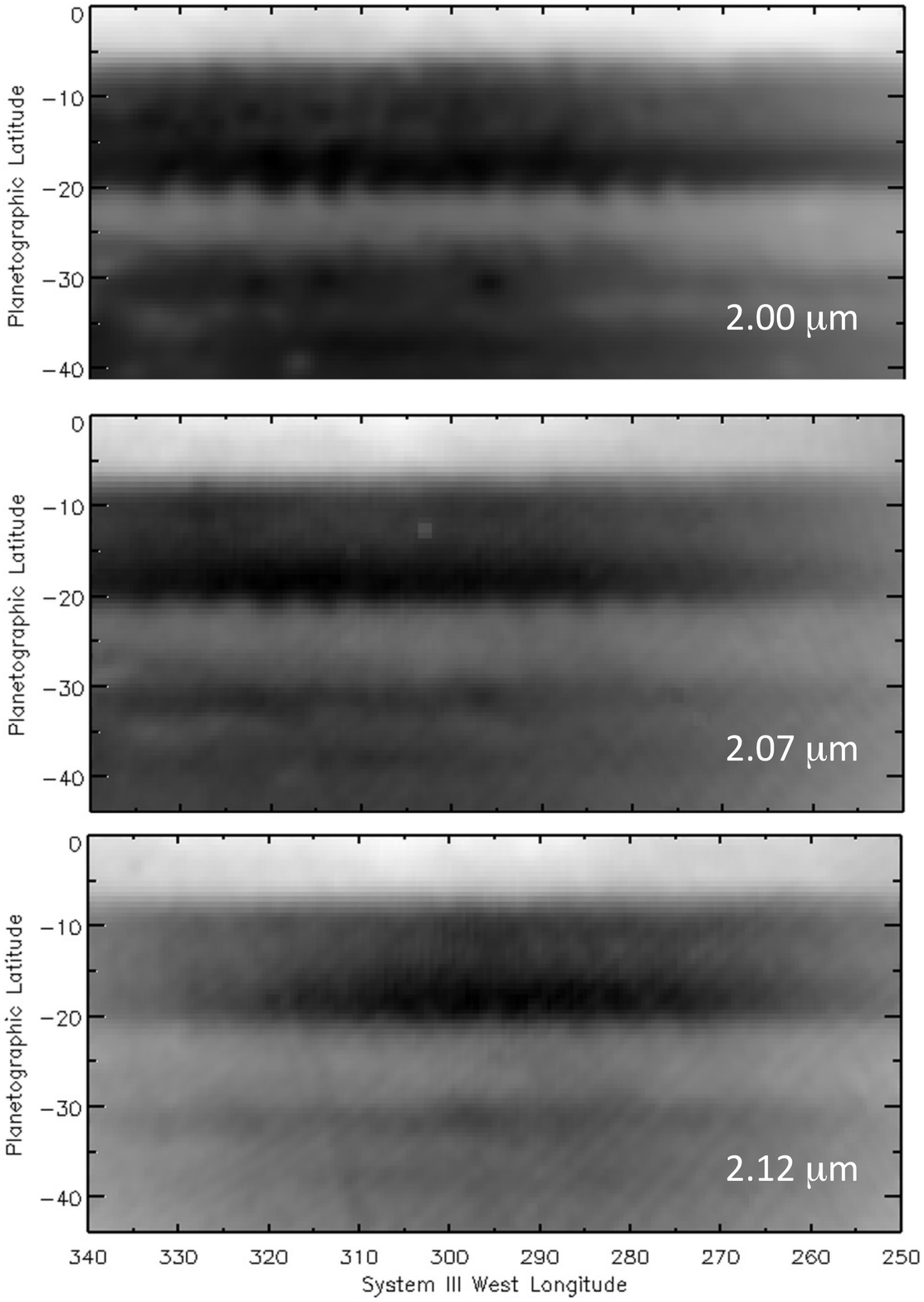}}
\caption{Maps of near-infrared reflectivity of Jupiter's disk, derived from images of Jupiter obtained on Oct 2, 2010, using the NSFCam2 facility instrument at the NASA Infrared Telescope Facility \citep{94sure} on a night of particularly good seeing.    The maps have been corrected for centre-to-limb behaviour, modified by a power law that allows faint features to be more visible and scaled between the minimum and maximum pixel values.   The wavelength of each map is identified in the lower-right corner. The sequence of maps is arranged from top to bottom in the order of sensitivity to particle reflectivity at increasing altitudes.  The faint cross-hatching in the 2.12-$\mu$m map is due to artefacts in the original image.  The different wavelengths correspond to different positions on a circular-variable filter (CVF) wheel with an approximately triangular transmission function and a full-width half maximum that is equivalent to a spectral resolution of ~1.8\%.  The SEBs wave at $20^\circ$S latitude decreases in visibility with increasing altitude.  The images from which these maps were made were acquired in IRTF program 2010B-010. }
\label{nir}
\end{centering}
\end{figure}

The near-infrared spectrum of Jupiter is dominated by sunlight reflected from atmospheric particulates.  The undulations of cloud opacity at the southern boundary of the SEBs are apparent at wavelengths sensing the tropospheric cloud decks in the 500-900 mbar region (Fig. \ref{nir}).  Figure \ref{nir} illustrates maps of cloud reflectivity taken from images in the middle of the SEB fade sequence on October 2, 2010 using the NSFCam2 facility instrument at the NASA Infrared Telescope Facility \citep{94sure}.   Observations at similar wavelengths also revealed the SEBs undulations near $20^\circ$S that are contemporaneous with the mid-infrared VISIR observations in November 2010 (Fig. \ref{visir}).

Going from 2.00 $\mu$m to 2.12 $\mu$m in Fig. \ref{nir}, the gaseous opacity of the atmosphere increases due to increasing absorption contributions from methane (CH$_4$) and molecular hydrogen (H$_2$).   The pressure corresponding to unit optical depth for nadir observations is approximately 900 mbar at 2.00 $\mu$m, 750 mbar at 2.07 $\mu$m and 500 mbar at 2.12 $\mu$m \citep[see Fig. 1 of][calculated for an aerosol-free atmosphere]{10sromovsky}.   Aerosols in the atmosphere will generally reduce the pressure values associated with these observations.  A full radiative transfer study is needed to identify the pressure range associated with these particles, but such a study is beyond the scope of this discovery paper.  We selected this wavelength range because of its rapid rise in gaseous opacity over a short spectral interval that appeared to be less sensitive to spectral variations of particulate albedo than shorter wavelengths.   It is clear that the undulations in the particulate boundary at $20^\circ$S that are evident at 2.00 $\mu$m are less distinct going from that wavelength to 2.07 $\mu$m and 2.12 $\mu$m.  These results are qualitatively consistent with the conclusions from the mid-infrared observations that the characteristic vertical extent of the wave is on the order of a scale height, a factor of $\sim2.7$ in pressure or $\sim20$ km at 500 mbar.

\section{Results: Parameters of the wave pattern}
\label{results}

\begin{figure*}[!htb]
\begin{centering}
\centerline{\includegraphics[angle=0,scale=0.6]{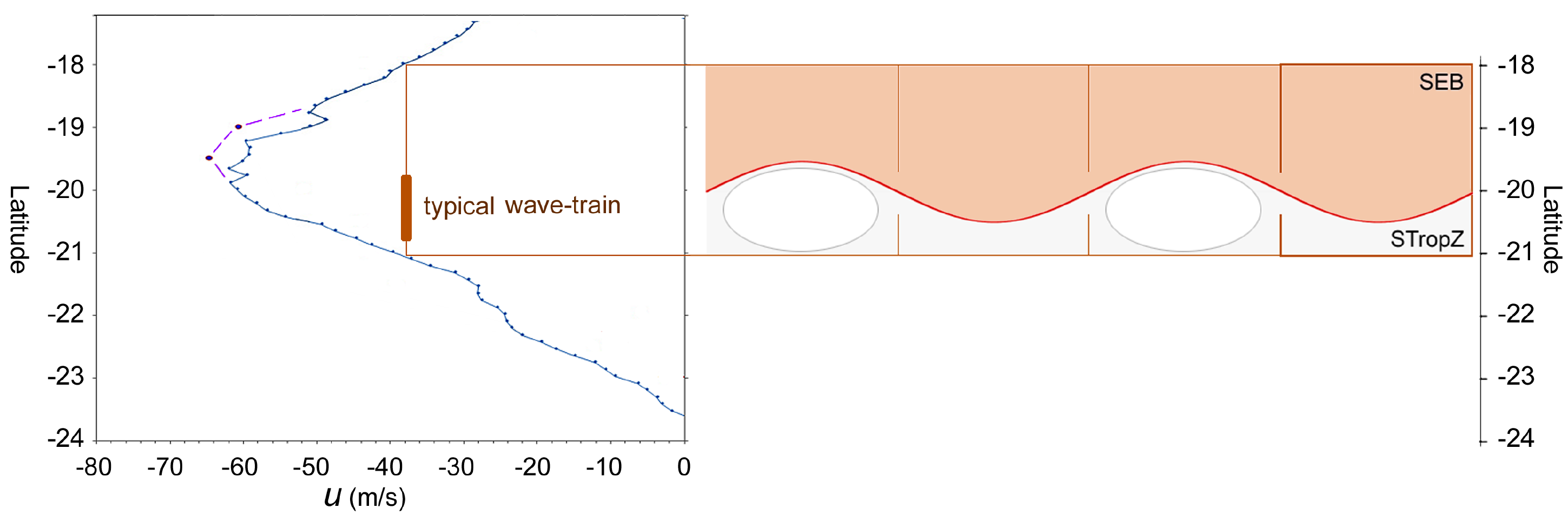}}
\caption{Cartoon comparing the general appearance of the wave pattern and associated white ovals with the zonal wind profile (ZWP) measured by Cassini \citep[blue line][]{03porco}.  The dashed line and two purple points is the ZWP indicated by New Horizons \citep{08cheng}.  On the diagram of the wave train separating the SEB and STropZ, the square outlines indicate that the wavelength (km) is approximately twice the width of the jet at the phase speed of the wave pattern (Fig. \ref{correlation}).  Latitude measurements are subject to an estimated uncertainty of $\pm0.3^\circ$.}
\label{cartoon}
\end{centering}
\end{figure*}

Both the reflectivity observations and the thermal-infrared observations revealed the presence of a wave pattern on the SEBs, with a phase speed that differs from the known wind speed at that latitude.  Intriguingly, both the wavelength and the speed have varied considerably from one year to another, and more modestly from one longitude sector to another.  Fig. \ref{cartoon} shows the latitudinal location of notable features associated with the SEBs wave pattern to provide guidance for the following discussion.

\subsection{Comparison to wind field}

\begin{figure*}
\begin{centering}
\centerline{\includegraphics[angle=0,scale=0.7]{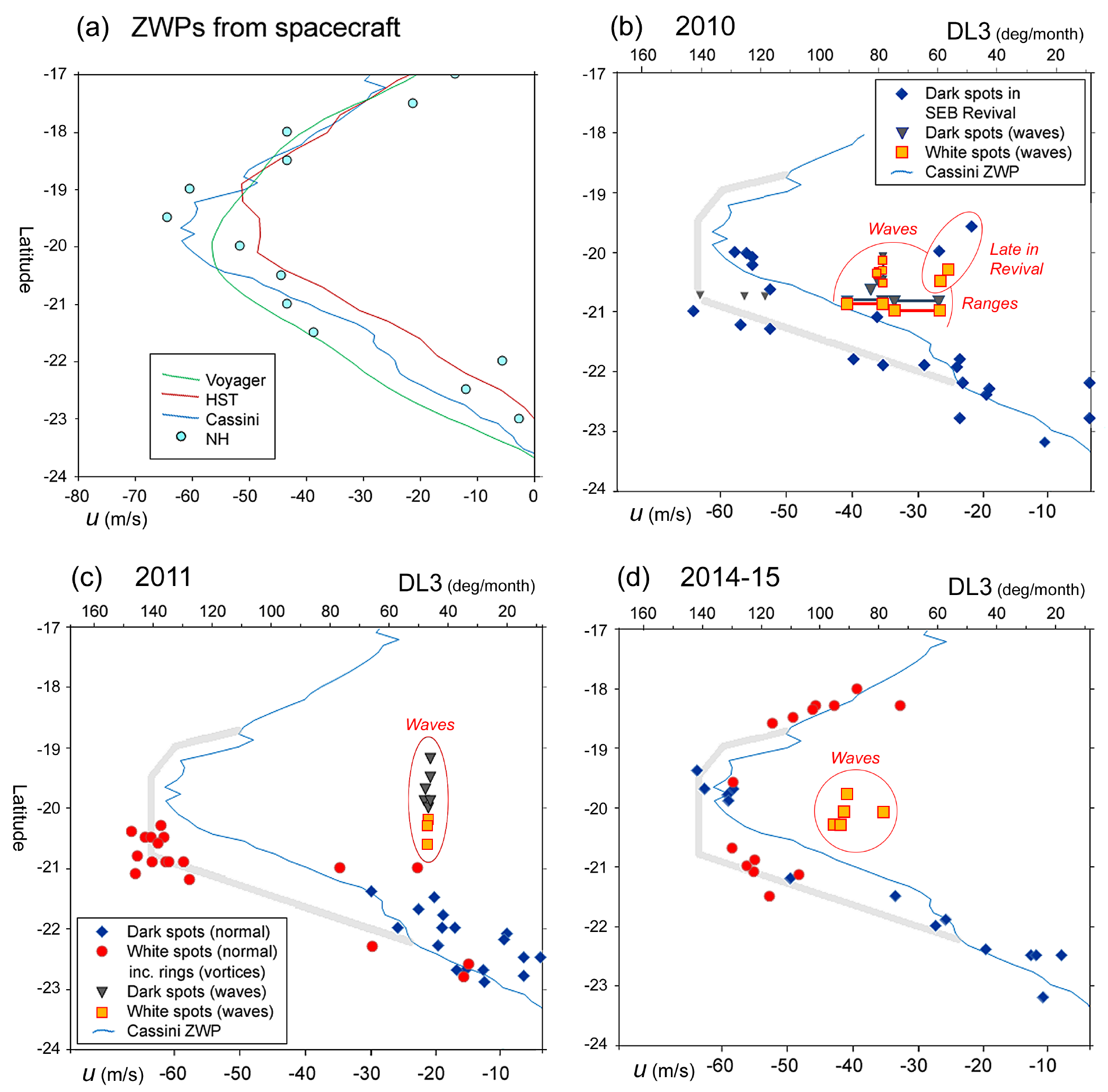}}
\caption{Comparison of spacecraft zonal wind profiles (ZWPs) (panel a) with measured drift rates of spots and waves in each epoch (b-d).  Panel a shows ZWPs from four spacecraft.  Voyager 1 \citep[green line,][]{86limaye}: the peak speed was that of the numerous vortices then present, but these may have masked a more rapid true jet peak. Hubble Space Telescope \citep[red line,][]{01garcia}: an average of ZWPs on four different dates between 1995 and 1998. Cassini \citep[blue line,][]{03porco} and New Horizons \citep[blue circles,][]{08cheng}: the fastest retrograding zonal windspeed is more consistent with the subsequent ground-based observations.  Panels b, c, and d compare the Cassini ZWP (blue line) to features analysed in the ground-based data:  waves (squares and triangles), other dark spots (diamonds) and white spots and vortices (circles).  The wave features are enclosed in ellipses or circles. Both zonal windspeed and DL3 (degrees per 30 days in System III) are shown.  Typical uncertainties are 1 m/s for speeds and $0.3-0.5^\circ$ latitude for positions, as explained in Section \ref{vis}.  Panel b shows the first epoch, July to November 2010, wave features during the SEB Fade (averages for different sectors, enclosed in a semicircle), and the second epoch, to January 2011 during the SEB revival (all other points, including wave features enclosed by an oval which were reappearing as the revival proceeded). Panel c shows the third epoch, June-September 2011 (wave phase speeds are for individual spots within a single wave train). Panel d shows the fourth epoch, October 2014 to March 2015 (wave phase speeds are averages for separate wave trains).  The grey line in panels b-d is a hypothetical SEBs jet profile, broader than that of Cassini, but nevertheless more consistent with the ground-based feature tracking of many dark spots (diamonds) and vortices (circles).}
\label{zwp_compare}
\end{centering}
\end{figure*}


The measurements of speed and latitude for all spots in the region are compared to zonal wind profiles (ZWP) in Fig. \ref{zwp_compare} to show the differences between the observed drift rates and the background zonal winds.  This shows that most of the discrete bright and dark spots observed in the visible data follow a consistent zonal drift profile which agrees with the Cassini ZWP (blue line in all panels), with the exception of a collection of fast-moving spots that tend to lie $\sim1^\circ$ south of the jet peak of the nominal ZWP between $20-22^\circ$S. These mostly appeared to be vortices (from their oval form and latitude, shown as circles in Figs. \ref{zwp_compare}(b)-(d)), although they also included irregular dark spots (shown as diamonds) during the SEB Revival.  The mismatch between their speed and the ZWP could be due to a dynamical property of the vortices \citep{08legarreta}.  However, we note that the observations appear to suggest a broader, faster jet profile than indicated by the Cassini ZWP (the grey line in Figs. \ref{zwp_compare}(b)-(d)), sensed by the discrete white and dark spots that are unrelated to the SEBs wave.   The vortices observed by Cassini fit onto this more southerly profile, and so do many spots in our zonal drift profiles for 2005-2015, both in normal times, and during the SEB Revivals in 2007 and 2010.   So does the peak of the jet as recorded by New Horizons \citep[$u=-64.5$ m/s, at $19.5^\circ$S][]{08cheng}, although those data have large scatter.  We speculate that these features may suggest the presence of a broader SEBs jet below the cloud-tops.  However, further discussion of this issue would be beyond the scope of this paper and we adopt the Cassini ZWP for the remainder of this article.  

Nevertheless, Fig. \ref{zwp_compare} shows that the wave features (squares and diamonds) are entirely distinct from those of the background flow, with slower retrograding speeds spanning the latitudes of the jet peak.  Note that the spread in latitude is largely due to measurements of small white spots and dark ÔprojectionsÕ on the south side of the wave pattern, and protruding slightly beyond the sinusoidal wave itself. All the data are consistent with the wave itself having a fixed central latitude between $19.5-20.0^\circ$S, exactly on the peak of the jet in Fig. \ref{zwp_compare}.  The white spots indicated by Fig. \ref{cartoon} were observed on the southern flank of this wave, down to $21^\circ$S.
 
The altitude of the wave pattern is not strongly constrained from the amateur imaging.  Its behaviour during the SEB Revival could suggest that its white ovals lay above the rapidly-retrograding dark patches.  On the other hand, its appearance in various wavebands suggests that it is intrinsic to the jet as observed at cloud-top level: thus it demarcates the visible SEBs edge, both in visible and thermal-IR images; and it visibly distorts passing vortices. In Section \ref{discuss} we show that the background atmospheric conditions (thermal profiles, wind profiles, etc.) are more consistent with the SEBs wave as a cloud-level feature than with one at high altitude.

\subsection{SEBs wave properties}

\begin{figure}
\begin{centering}
\includegraphics[angle=0,scale=0.9]{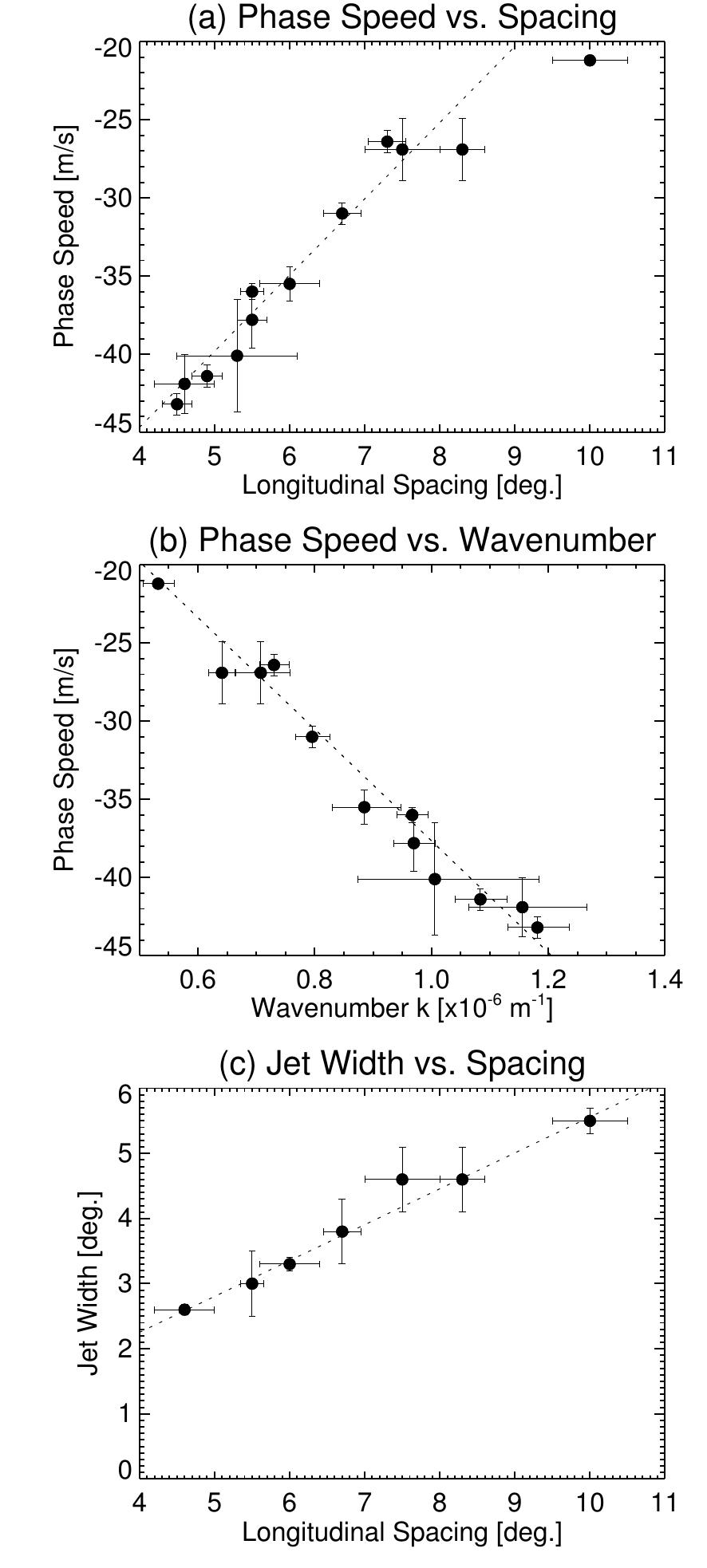}
\caption{Relationship between the phase speed $c_x$ (i.e., negative with respect to System III) and longitudinal spacing of the SEBs wave shown in panel a, with a linear trendline (dashed line).   Panel b re-expresses the longitudinal spacing in terms of a horizontal wavenumber $k$. Panel c shows the correlation between the width of the ZWP at the phase speed of the wave and the longitudinal spacing. }
\label{correlation}
\end{centering}
\end{figure}

Having established that the SEBs wave was retrograding more slowly than the jet peak (i.e., the phase speed $c_x$ is negative but slower than the retrograding zonal windspeed $u$), we now use the phase speed and wavelength measurements from all 2010-2015 epochs (Table \ref{waveparams}) to understand the properties of the wave in Fig. \ref{correlation}, utilising drifts and locations calculated via the JUPOS project.  Fig. \ref{correlation}a shows that we find a strong correlation between the phase speed of the wave and the longitudinal spacing (wavelength, $\Delta\phi$), with a linear regression of $c_x=(4.9\pm0.4)\Delta\phi-(64.2\pm2.2)$ m/s.  This implies that smaller wavelengths are retrograding with a faster (more negative) phase speed.  It may be significant that the phase speed extrapolates to $c_x=-(64.2\pm2.2)$ m/s for zero spacing - very close to the peak jet speed.  However, we note that the longest wavelength of the SEBs wave pattern ($\Delta\phi\approx10^\circ$) deviates from this simple linear trend.  Nevertheless, such a relationship strongly indicates that this is a modestly retrograding wave train on the peak of the SEBs jet that can exist in a variety of conditions of the SEB (normal, faded, or energetically reviving).  

To aid assessment of the dispersion relation for this wave pattern, Fig. \ref{correlation}b re-expresses the correlation in terms of the zonal wavenumber $k$, using a distance scale of approximately 1181 km/degree-longitude (exact value depending on the observed latitude of the wave).  We find a linear regression of $c_x=-(35.7\pm2.4)\times10^{6}k-(1.9\pm2.0)$ m/s reproduces the observed data within the $1\sigma$ uncertainties.  As a corollary of this correlation, the frequency of the wave pattern relative to the peak flow of the jet is almost invariant, with a mean of $(4.09\pm0.22)\times10^{-6}$ s$^{-1}$, equivalent to a period of $2.83\pm0.15$ days, which is approximately the time taken for full-speed vortices to travel along one wavelength of the wave-train.  We also investigated whether the data exhibited a relationship of the form $u-c_x\propto1/k^2$ (typical of planetary wave dispersion relationships, see Section \ref{discuss}), and found that a relation $(u-c_x)=-(11.9\pm1.0)\times10^{-12}/k^2 - (11.5\pm1.2)$ m/s could reproduce the data for large values of $k$, but not the smallest (i.e., this relation fails to reproduce longitudinal wavelengths exceeding $8^\circ$).  We explore the implications of these correlations between phase speed and wavelength in Section \ref{discuss}.

There is no correlation between the measured phase speed and the latitude (Fig. \ref{zwp_compare}).  The typical latitudinal amplitudes of the waves (north to south, peak to peak) in the images were $1.2\pm0.3^\circ$ in November 2010 (measured from the 8.6 $\mu$m image), $1.6\pm0.5^\circ$ in January 2011, $1.4\pm0.3^\circ$ in August 2011 and $0.8\pm0.3^\circ$ in February 2015.  This latitudinal amplitude is $14-21$\% of the longitudinal wavelength. The small white ovals that occupied troughs on the south side of the SEBs wave had meridional widths $1-2\times$ the wave amplitudes.  The observations of the SEBs wave pattern therefore reveal latitudinal amplitudes smaller than $2^\circ$ and a vertical confinement to the region of the cloud tops.  

To investigate this latitudinal confinement in more detail, we measure the latitudinal width of the ZWP at the phase speed of the wave, and define this as the 'jet width.'   If we plot the jet width versus the longitudinal spacing, we find that there is an exact proportionality in Fig. \ref{correlation}c - the latitudinal width ($\Delta\psi$) of the waveguide is almost exactly half of the longitudinal wavelength ($\Delta\phi$), with a linear regression of $\Delta\psi=(0.55\pm0.07)\Delta\phi$.  Converting from degrees to kilometres in the latitudinal ($\Delta y$) and longitudinal ($\Delta x$) directions, this relationship becomes $\Delta y=(0.52\pm0.07)\Delta x$. Smaller longitudinal spacings (i.e., larger zonal wavenumber, $k$) imply smaller latitudinal widths (i.e., larger meridional wavenumber, $l$) for this type of wave.  In the next section, we use these correlations between the phase speed, meridional and zonal wavenumbers to understand the dynamics of the SEBs wave pattern.

%

\begin{table*}[htdp]
\caption{Parameters of the SEBs wave pattern for the different epochs.  Where dates are repeated over two rows, this implies separate wave trains with different spacings/speeds.  The latitudes (planetographic) are for the white ovals in the troughs of the wave, the wave pattern itself is at $19.5-20.0^\circ$S.  $N$ is the number of spots in the wave group(s) that showed the most uniform spacing and speed; in some cases more spots were present with more variable spacing or speed.}
\centering
\begin{tabular}{|c|c|c|c|c|c|}
\hline
Year/SEB State	& Month & Phase Speed (m/s) & Lon. Spacing & Central Latitude & N Spots \\
\hline
2010 & July & $-40.1\pm3.6$ & $5.3\pm0.8$ & $20.8\pm0.5^\circ$S &  11 \\
(SEB faded) & July-Sep &$-37.8\pm1.8$ &	$5.5\pm0.20$ &	$20.9\pm0.5^\circ$S &	37 \\
 & Aug-Sep &$-26.4\pm0.7$ &	$7.3\pm0.25$ &	$20.8\pm0.5^\circ$S &	14 \\
 & Sep &$-31.0\pm0.7$ &	$6.7\pm0.25$ &	$20.8\pm0.5^\circ$S &	7 \\
 & Nov &$-36.0\pm0.5$ &	$5.5\pm0.15$ &	$20.4\pm0.5^\circ$S &	8 \\
\hline
2010-11& Jan & $-26.9\pm2.0$ & $7.5\pm0.5$ & $20.0\pm0.5^\circ$S &	3 \\
(SEB revival) & Jan & $-26.9\pm2.0$ & $8.3\pm0.3$ & $20.5\pm0.5^\circ$S & 	7 \\
\hline
2011& Aug & $-21.2\pm0.3$ & $10.0\pm0.5$ & $20.4\pm0.2^\circ$S& 6 \\
(After Revival) & & & & & \\
\hline
2015 & Jan & $-43.2\pm0.7$ & $4.5\pm0.2$ & $20.3\pm0.2^\circ$S & 7 \\
(SEB Normal) & Feb & $-41.4\pm0.7$ & $4.9\pm0.2$ & $20.1\pm0.3^\circ$S & 6 \\
& Feb-Mar &$-35.5\pm1.1$ & $6.0\pm0.4$ & $20.1\pm0.1^\circ$S & 5 \\
& Feb-Mar & $-41.9\pm1.9$ & $4.6\pm0.4$ & $20.3\pm0.1^\circ$S & 9 \\
\hline
\end{tabular}
\label{waveparams}
\end{table*}%

\section{Discussion}
\label{discuss}

Besides the axisymmetric banded structure and plethora of discrete vortices present in Jupiter's visible atmosphere, wave activity constitutes an important process shaping both the cloud decks and the thermal structure of the atmosphere.  In Section \ref{results}, we reported the regular presence of a wave pattern on the edge of the SEBs (Jupiter's fastest retrograding jet) whose phase speed $c_x$ is linearly related to its wavelength (Fig. \ref{correlation}) and that is retrograding more slowly than the jet peak wind speed.  We investigate whether this clearly-defined dispersion relationship can provide insights into the nature of the wave.

\subsection{Rossby Waves}

It was initially noted that the SEBs wave bears a superficial resemblance to a Rossby wave - large-scale, low-frequency horizontal waves that owe their existence to the changing Coriolis force with latitude acting as a restoring force.  Although their dispersive properties show similar relationships to those found in Section \ref{results}, we show below that the SEBs wave cannot be a Rossby wave in the classical sense.  Rossby wave activity is evident on many of Jupiter's zonal jets as reviewed in Section \ref{intro}.  The best-established example comprises the well-known large dark formations, also known as infrared hot spots, on the prograding jet at $7^\circ$N (the southern edge of the North Equatorial Belt, NEBs).  There is considerable evidence that they represent horizontally-trapped Rossby waves, confined between the equator and the NEBs jet, whose prograde zonal windspeed is much faster than the observed phase speed of the dark formations \citep{90allison, 95dowling_SL9, 98ortiz, 00showman, 05friedson, 11garcia_jup}.  Accordingly, these Rossby waves on the NEBs prograde jet also show a direct relation between phase speed and wavenumber \citep{98ortiz, 06arregi, 08rogers_jup07, 08rogers_jup01_ptI, 08rogers_jup01_ptII}.

Here we focus on the retrograde jets at $17^\circ$N and $20^\circ$S (the NEBn and SEBs, respectively).  During the Cassini encounter in 2000, the NEB hosted a large-scale wave pattern that has been interpreted as a Rossby wave, although it had very different characteristics to the NEBs wave discussed above.  It was most clearly observed in late 2000 by ground-based images and by Cassini, and extended all around the NEB with a wavelength of $20-25^\circ$ longitude.  It was detectable as a diffuse variation of haze thickness and upper tropospheric temperature, only detectable above the main cloud deck, in methane-band and ultraviolet images and in thermal maps from the Cassini/CIRS \citep{03porco, 04flasar_jup, 04rogers, 06li}.  Following the analysis of earlier observations of slowly-moving thermal waves \citep{94orton, 97deming}, \citet{06li} modelled these as Rossby waves that were generated near the 0.5 bar level (near the visible cloud tops) and able to propagate vertically.  The phase speed ranged from $c_x=-3.9$ to $-1.9$ m/s \citep{04rogers}, intermediate between the retrograding NEBn speed at 700 mbar \citep[measured in continuum imaging by][]{98banfield} and the weakly prograding NEBn speed at 350-500 mbar \citep[measured in UV imaging by][]{06li_wind}.  Observations showed that the waves were locked to the positions of underlying tropospheric features where there was visible anticyclonic eddying, and anomalies in the wave motion were related to visible cloud events, implying that the waves were controlled by the underlying tropospheric wind patterns \citep{04rogers}.  These patterns probably constituted a large-scale horizontal wave along the NEBn retrograde jet, since the jet is thought to meander between cyclonic and anticyclonic circulations on either side of it. This motion is indicated in the Voyager and Cassini animated maps \citep[see Cassini movie: NASA/JPL/SWRI/CICLOPS, at \mbox{http://photojournal.jpl.nasa.gov}, PIA02863; ][]{03porco}, although it has yet to be described quantitatively and the animations also show much smaller-scale chaotic motion.  Thus the basis of the NEBn wave pattern may have been a Rossby wave \citep[as proposed by][]{97deming}, but with a much longer wavelength than what we have observed on the SEBs, and a near-zero phase speed $c_x$ that was less than the positive zonal windspeed $u$ in the 350-500 mbar regime (i.e., $u-c_x>0$).  

\begin{figure*}
\begin{centering}
\includegraphics[angle=0,scale=0.8]{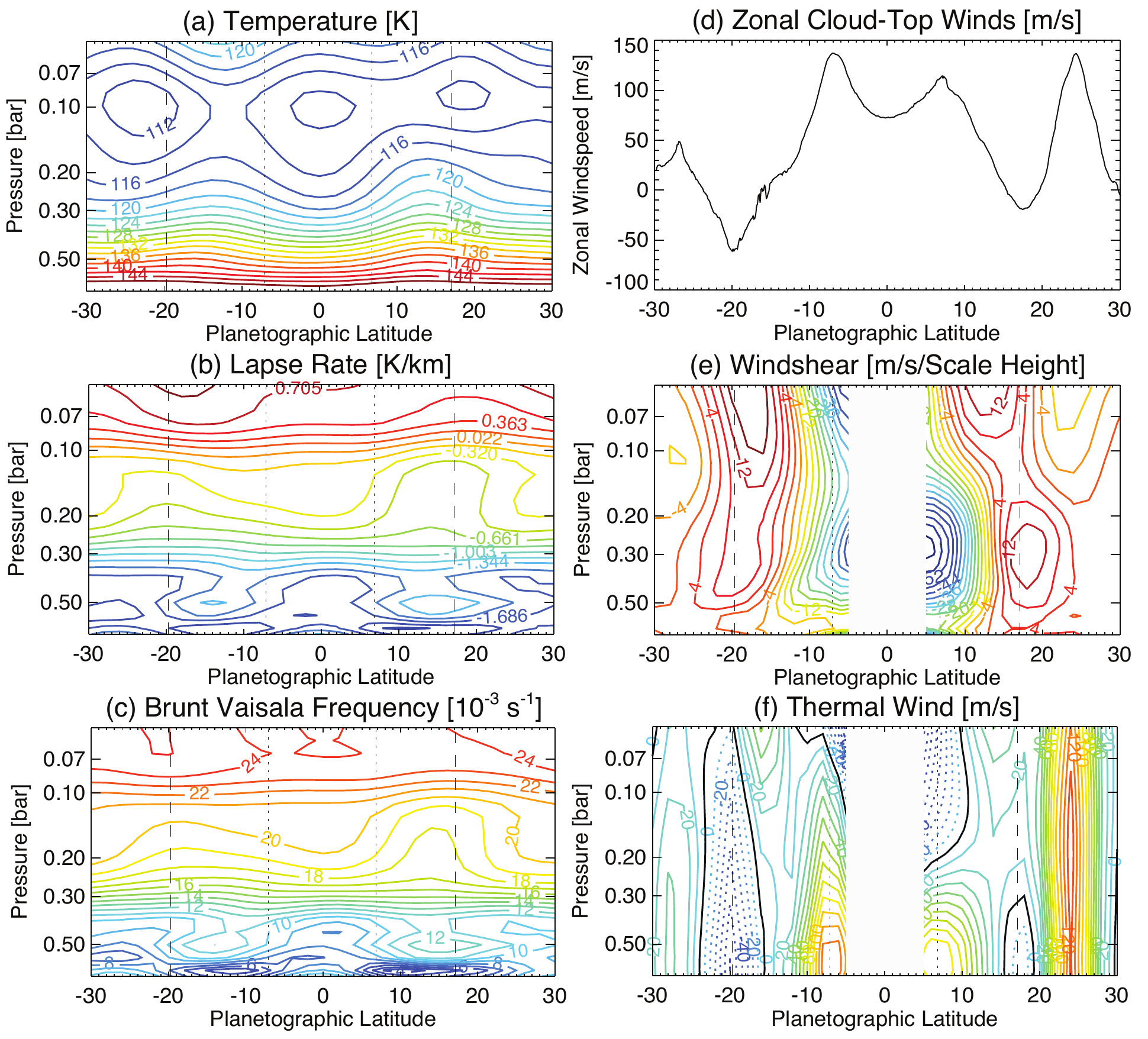}
\caption{Contrasting the temperature and wind fields associated with the SEBs and NEBn retrograding jets (vertical dashed lines) and the SEBn and NEBs prograding jets (vertical dotted lines).  Panel (a) shows the temperature cross-section derived from Cassini/CIRS \citep{09fletcher_ph3}, (b) shows the lapse rate $dT/dz$ and (c) provides an estimate of the buoyancy frequency $N$ based on the assumption of a frozen equilibrium for ortho- and para-H$_2$. The Cassini/ISS cloud-tracked zonal winds \citep{03porco} are in panel (d).  These are compared to the vertical windshear per scale height in panel (e).  This is used to estimate zonal mean winds at all altitudes in panel (f),  where negative speeds are denoted as dotted contours and positive speeds are solid contours (contour spacing 10 m/s).  Here solid black lines indicate regions where $u=0$ m/s.}
\label{thermal}
\end{centering}
\end{figure*}

However, the newly-discovered SEBs wave is propagating in a very different background environment compared to the NEBn wave.  To investigate this background, we used Jupiter's temperature structure derived from Cassini Composite Infrared Spectrometer (CIRS) mapping acquired during the December 2000 flyby \citep{04flasar_jup, 06simon, 09fletcher_ph3}.  Specifically, we used the spectral inversions of zonal mean spectra from the ATMOS02A map (December 31, 2000) by \citet{09fletcher_ph3}, which used the NEMESIS optimal estimation retrieval algorithm \citep{08irwin}.  Given that Jupiter's atmosphere is highly variable, and the thermal structure may have altered significantly since the time of the Cassini flyby, the properties of the background atmosphere are meant only as a qualitative guide.  These background properties are shown for the equatorial region surrounding the four jets of interest in Fig. \ref{thermal}, revealing the temperature structure $T(z,\psi)$ at each altitude and latitude ($\psi$), the measured lapse rate $\Gamma=-\partial T/\partial z$, atmospheric scale height ($H=RT/g$, where $R$ is the molar gas constant divided by the mean molar weight of Jupiter's atmosphere and $g(\psi,z)$ is the gravitational acceleration) and buoyancy frequency (described below), compared to Cassini/ISS measurements of the zonal wind speeds.  Uncertainties in the derived temperatures range from 1.9 K at the tropopause to 1.0 K at 500 mbar and 1.7 K at 1 bar, although these will become magnified when derivatives are taken to calculate winds and vorticity gradients with altitude and latitude.

\begin{figure*}
\begin{centering}
\centerline{\includegraphics[angle=0,scale=0.75]{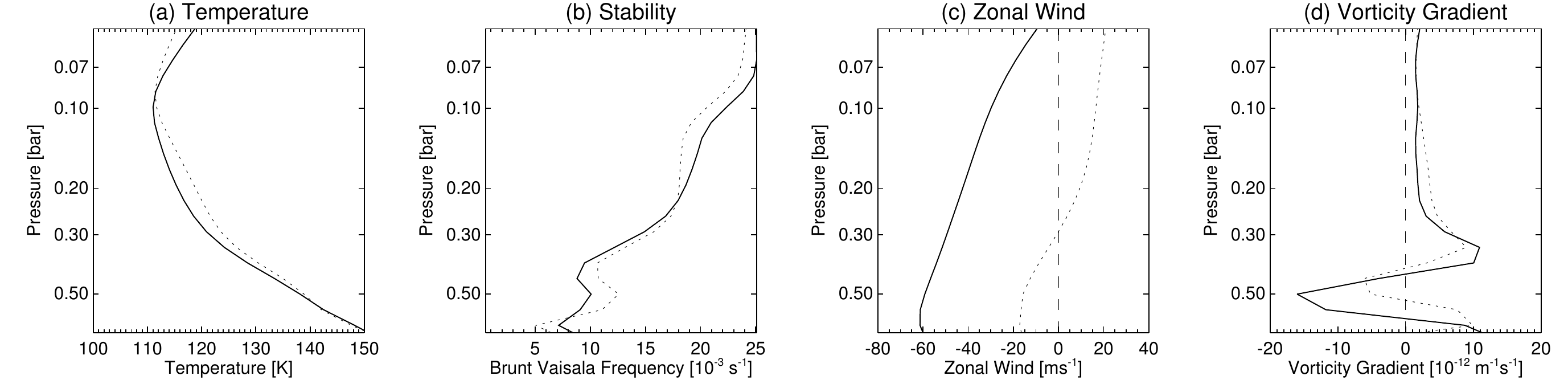}}
\caption{Comparing the temperature profile, vertical stratification (the Brunt Vaisala frequency), zonal wind (extrapolated via the thermal wind equation) and `effective $\beta_e$' (the vorticity gradient) as a function of altitude for the two retrograding jets:  the SEBs (solid line, $20^\circ$S) and the NEBn (dotted line, $17^\circ$N).  The vertical dashed line indicates the zero point.}
\label{profiles}
\end{centering}
\end{figure*}

The gradients in the temperature structure allow us to derive the zonal windspeed $u(\psi,z)$ as a function of altitude and latitude using the thermal wind equation in the zonal direction \citep{87andrews}:
\begin{equation}
f\pderiv{u}{\ln \left( p \right)} = \frac{R}{a} \pderiv{T}{\psi} = R\pderiv{T}{y} = -fH\pderiv{u}{z}
\end{equation}
where $a(\psi)$ is the planetary radius, $f=2\Omega \sin(\psi)$ is the Coriolis parameter that depends on the planetary angular velocity ($\Omega$) and latitude $\psi$, and all other terms have been previously defined.  We use the Cassini-derived windspeeds of \citet{03porco} and assume the cloud-tracked tracers reside at 500 mbar, then integrate with altitude to determine the zonal thermal winds in the domain of interest.  These winds are shown in Fig. \ref{thermal}f  and assume a geostrophic balance between the temperature gradients and the vertical shears on the zonal jets.  The temperatures, stability and zonal wind are compared for both the SEBs and NEBn in Fig. \ref{profiles}a-c.  The retrograding speeds of both the SEBs and NEBn jets decline with altitude, but the strength of the windshear shows a stark asymmetry between the two jets (Fig. \ref{thermal}e). Whereas the NEBn jet becomes prograde for $p<300$ mbar \citep[qualitatively consistent with the findings of][]{06li}, the SEBs jet remains retrograde throughout the region constrained by the thermal data (80-700 mbar).  Unfortunately this windshear estimate cannot be verified because of the difficulty in assigning precise altitudes to cloud tracers to assess what region they reside in, but the inferred retrograding windspeed has stark consequences for the SEBs wave.

The dispersion relationship for a 3D Rossby wave propagating in a baroclinic atmosphere \citep{87andrews,96achterberg, 06li, 11sanchez_book} is given by:
\begin{equation}\label{eq:dispersion}
u-c_x = \frac{\beta_e}{k^2 + l^2 + (f^2/N^2)(m^2+n^2)}
\end{equation}
where $u$ is the zonal windspeed of the background flow; $c_x$ is the phase speed of the observed wave (both relative to System III); $k$, $l$ and $m$ are the zonal, meridional and vertical wavenumbers of the wave.  This relationship also depends on the vertical stratification via the buoyancy frequency ($N$, also known as the \textit{Brunt V\"{a}is\"{a}l\"{a}} frequency), where:
\begin{equation}
N^2=\frac{g}{T}\left(\pderiv{T}{z} + \frac{g}{c_p}\right)
\end{equation}
Here $c_p$ is the specific heat capacity assuming a equilibrium between the two spin isomers of hydrogen (ortho and para-H$_2$), and includes contributions from latitudinally-invariant helium and methane.  The zonal mean temperatures derived from CIRS data were used to calculate the buoyancy frequency shown in Fig. \ref{thermal}c, revealing that the atmosphere remains stably stratified throughout the altitude region of interest (the observable clouds are expected to reside at the 400-600 mbar level), only becoming unstable to convection for pressures exceeding 700 mbar.  The warm SEB and NEB exhibit stronger stratification than the cooler equatorial zone, although the vicinity of the SEB is warmer and more stable than the NEB in the 100-200 mbar range, at least during the Cassini epoch.  All regions show a rapid increase of $N$ with altitude into the stable stratosphere.  

The final component of the Rossby wave dispersion relation is the restoring force imposed by the latitudinal gradient of quasi-geostrophic potential vorticity (QGPV, $q_G$), known as the `effective beta' $\beta_e=\partial{q_G}/\partial{y}$, \citep[e.g., p127 of][]{87andrews}.  $q_G$ is conserved by these flows on Jupiter due to the small Rossby numbers (the influence of inertial forces is smaller than those of Coriolis forces) associated with this wave pattern ($Ro=U/fL<0.1$ for the characteristic speeds $U$ and longitudinal spacings $L$ of this wave).   Crucially, a positive value of $\beta_e$ provides a stable restoring force for latitudinal excursions that can permit the presence of Rossby waves.  The zonal winds shown in Fig. \ref{thermal}f show that the SEBs wave has $u-c_x<0$ (i.e., the waves phase moves east with respect to the zonal wind direction), which cannot occur if all the terms in Eq. \ref{eq:dispersion} are positive.  Conversely, the NEBn wave has $u-c_x>0$, which is perfectly acceptable.  Although $\beta_e$ can be negative (as we discuss below), this would no longer provide a restoring force for latitudinal excursions, so the SEBs wave cannot be interpreted as a Rossby wave in the classical sense, despite the observed morphological similarities and the $u-c_x\propto1/k^2$ relation described in Section \ref{results}.

\subsection{Instabilities}

Oscillations observed in planetary atmospheres can be due to either mechanical forcing or instabilities of some form.  The dimensionless Richardson number $Ri=N^2H^2/(\pderiv{u}{\ln(p)})$ measures the importance of the atmospheric stratification against vertical shears on the zonal winds, and we find this to be positive (as the atmosphere is statically stable, $N^2>0$) and large ($Ri>>1$) throughout the 0.1-1.0 bar region.  This rules out Kelvin-Helmholtz instabilities ($0<Ri<0.25$) due to strong vertical wind shears, but permits other types of instabilities (barotropic and baroclinic) in the region of interest, which might generate waves.

\begin{figure}
\begin{centering}
\includegraphics[angle=0,scale=0.8]{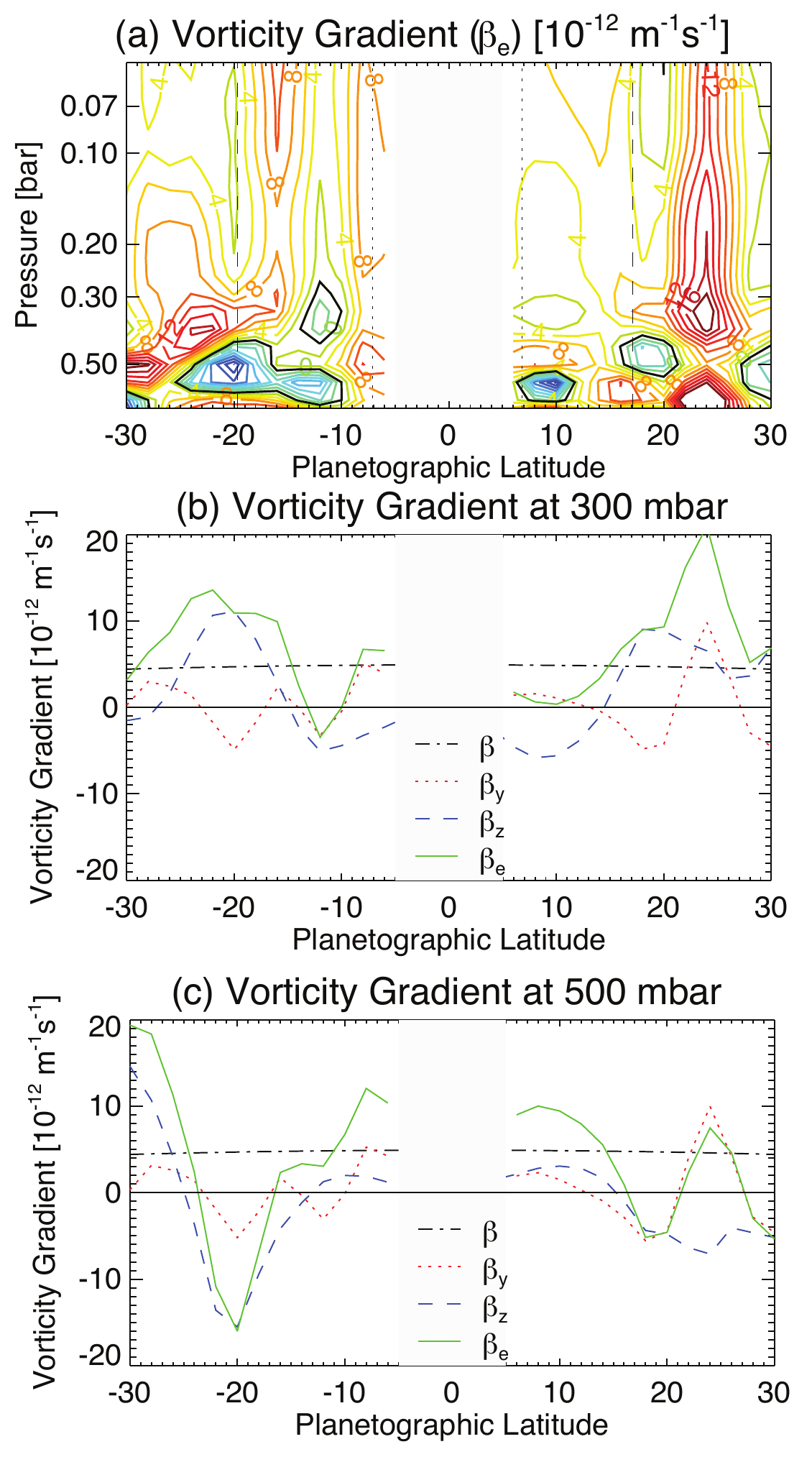}
\caption{The gradient of quasi-geostrophic potential vorticity, known as `effective beta' $\beta_e$, is plotted with pressure and latitude in panel (a).  Retrograde jets are shown as vertical dashed lines, prograde jets as dotted lines.  Regions enclosed by black contours exhibit negative $\beta_e$, a condition for instabilities to occur.  In panels (b) and (c) we show all of the contributions to $\beta_e$ at 300-mbar and 500-mbar, respectively, as described in the main text.  Panel (b) shows that $\beta_e$ is mostly positive at 300 mbar and reproduces the findings of \citet{06li} for the region of the NEB.  Panel (c) demonstrates that a region of negative $\beta_e$ is found near the SEBs at 500 mbar.}
\label{qgpv}
\end{centering}
\end{figure}

To investigate the stabilities of the SEBs and NEBn jets in more detail, we examine the latitudinal $q_G$ gradient in more detail.  Sign reversals for $\beta_e$ as a function of latitude are a necessary (but not sufficient) condition for baroclinic instabilities that might be a source of the observed wave phenomena on Jupiter and Saturn \citep{06read_jup, 09read}.  The effective beta parameter is the sum of $\beta=\partial{f}/\partial{y}$, the northward gradient of planetary vorticity (i.e., the Coriolis parameter); $\beta_y=-\partial^2u/\partial{y^2}$, the meridional curvature of the zonal wind field (the relative vorticity gradient); and $\beta_z$, the vertical curvature of the wind field (the baroclinic term).   Following \citet{87andrews, 11sanchez_book}:
\begin{eqnarray}
\beta_e=\beta+\beta_y+\beta_z \\
\beta_e=\beta-\frac{\partial^2u}{\partial y^2}-\frac{1}{\rho}\pderiv{}{z}\left(\rho \frac{f^2}{N^2}\pderiv{u}{z}\right) 
\end{eqnarray}
Here $\rho=p/RT$ is the atmospheric density from the ideal gas equation.   The Cassini-derived windspeed and thermal structure allow us to evaluate $\beta_e$ in Fig. \ref{qgpv}, showing that the baroclinic contribution to $\beta_e$ overwhelms the Coriolis gradient ($\beta$) in the vicinity of the cloud tops near the SEBs jet (by a factor of approximately $\sim2\beta$), and also near the NEBn jet (by a factor of $\sim1\beta$).   The vertical structure of $\beta_e$ is shown for the NEBn and SEBs jets in Fig. \ref{profiles}d, and is largely determined by the strong variability observed in the vertical stratification (e.g., the buoyancy frequency in Fig. \ref{thermal}).  This generates a region of strong negative $\beta_e$ near the SEBs, and such a sign change is a necessary (but not sufficient) condition for baroclinic instability following the Charney-Stern criterion \citep{62charney}.   Furthermore, if we ignore the baroclinic component (i.e., ignore variations of the windfield with altitude) and evaluate the latitudinal gradient of \textit{absolute} vorticity $\beta_e=\beta+\beta_y$, we find that $\beta_e<0$ \citep[previously shown by][]{06read_jup}, providing a necessary (but not sufficient) condition for barotropic instability according to the Rayleigh-Kuo criterion \citep{49kuo}.  Both of these stability criteria are therefore violated in the case of the SEBs jet at cloud level, but this does not guarantee the development of unstable modes.   

We used our estimates of $N$ and $H$ to calculate the internal Rossby deformation radius ($L_d=NH/f$, the natural scale of motions in the stably-stratified atmosphere) which varies between 1000-2000 km ($1-2^\circ$ longitude) near the cloud tops.  This is smaller than the observed $4-10^\circ$ longitude spacing of the SEBs wave, suggestive of a baroclinic instability mechanism generating the wave.  Furthermore, the Rhines scale \citep[$L_\beta=2\pi\sqrt(2U/\beta$),][]{73rhines}, which defines an approximate upper limit to the size of baroclinic instabilities, varies between 20,000-25,000 km (depending on the value assumed for the characteristic speed $U\sim c_x$), which is larger than the longest wavelength detected ($10^\circ$ longitude or $\sim11,000$ km).

If an instability were to generate unstable modes, then this could manifest as a meandering of the original jet, superficially resembling the SEBs wave pattern.  The baroclinic instability mechanism was also proposed as an explanation for Saturn's ribbon wave (on an eastward jet) and associated small eddies \citep{86godfrey}.  We note that the region where the Charney-Stern criterion is violated is close to the cloud tops, consistent with the observational conclusion that the wave pattern is not present in the upper troposphere.   We speculate that an initial instability near the cloud tops could generate a wave that would equilibrate at some scale to form the uniformly propagating wave that we have observed.  However, quantitative simulations of the growth of unstable modes is beyond the scope of the present analysis.

At lower pressures (300 mbar) we find that $\beta_e>0$ throughout Jupiter's low latitudes ($\pm30^\circ$), which is consistent with the northern-hemisphere calculations of $\beta_e$ from \citet{06li}, which were entirely positive at the location of the NEB wave at 315 mbar.  This implies a vertical transition from the baroclinically-unstable cloud tops to the more stable upper atmosphere, except that the NEBn wave was permitted to propagate vertically by the prograde zonal flow in the upper troposphere.  The existence of regions of positive $\beta_e$ overlying regions of negative $\beta_e$ suggests a complex baroclinic structure, which may have implications for the types of instabilities that are energetically favoured in this region \citep[e.g.,][]{81conrath_stab}.  A caveat to this discussion is the atmospheric conditions could certainly have altered between the observations of the SEBs waves (2010-2015) and the Cassini flyby in 2000.  Indeed, temperature gradients across the SEBs jet were observed to vary during the fade and revival cycle of the SEB \citep[][see their Fig. 6]{11fletcher_fade}, and the high-resolution images in Fig. \ref{visir} hint at thermal gradients near the SEBs jet that are stronger than those revealed by Cassini, although the uncertainties associated with temperatures derived from mid-IR photometric imaging are rather large \citep{09fletcher_imaging}.  Furthermore, the north-south temperature gradients would have to be longitude-specific rather than zonally-averaged (as they have been in this study), as the contrasts have been observed to vary with longitude, particularly during the vigorous revival sequence.  The evolution of these negative $\beta_e$ regions over time, particularly in association with belt/zone upheavals, will be the topic of a subsequent study.  

Finally, we note that Jupiter's SEBs wave pattern superficially resembles Saturn's `string of pearls', a chain of cyclonic vortices bright at 5-$\mu$m (and hence cloud-free) on a retrograding jet.  It was also detectable in near-infrared light.  However, unlike the SEBs, the wave appeared to be retrograding faster than the jet peak \citep{14sayanagi} so that $u-c_x>0$, and no relationship between $c_x$ and longitudinal wavelength was identified.  

In summary, the gradient reversals in QGPV (baroclinic) and absolute PV (barotropic) near the cloud-tops provide the necessary conditions for violation of baroclinic and barotropic stability criteria for the SEBs jet.  If unstable modes are able to grow, we speculate that they equilibrate to the meandering pattern observed in our visible and infrared datasets, with a dispersion relation with $c_x\propto-k$ and a maximum phase speed of $-64.2\pm2.2$ m/s as the longitudinal spacing tends to zero.  

\subsection{Variability of the SEBs wave}

Why has the SEBs wave pattern been observed only in these recent years?  It is not always present; indeed it seems to be rarely if at all present during years of `normal' activity in the SEB, being absent during the Voyager and Cassini encounters.  We speculate that it may develop when the SEB is generally quiet, lacking in large-scale convective activity (`rifting').  In summer 2010 (epoch 1), the wave pattern was present when the SEB was completely quiet during the Fade.  Inspection of the  few high resolution images available from previous SEB Fades also reveal that the SEB wave pattern was present during complete SEB Fades in 1962, 1974, 1990 and 1993.  The SEBs wave pattern was imaged by Pioneer 11 in December 1974 \citep{77rogers} and by the Observatoire du Pic du Midi in September 1962\footnote{Images available in the Base de Donn\'{e}es d'Images Plan\'{e}tairesÕ at: http://www.lesia.obspm.fr/BDIP/index.php}, January 1990 \citep[Figs. 5 and 6 of][]{92rogers} and April 1993 \citep[Fig. 4 of ][]{96sanchez_jup}.  However it was not visible in 2007 nor 2009, suggesting that it only appears when the SEB fading is complete. 

In August and September of 2011 (epoch 3) the SEB was also quiet, as the SEB Revival had finished and the usual rifting west of the GRS did not resume until late September.  In 2014-15 (epoch 4), the usual rifting was confined to a very small patch west of the GRS, and the wave pattern was confined to longitudes east of the GRS.  On the other hand, the SEB was not at all quiet in January 2011 (epoch 2), when the SEB Revival comprised enormous disturbances within the SEB and large, rapidly-changing dark spots were retrograding on the SEBs. The wave pattern was still detectable during this epoch, although it is possible that the reviving SEB had not yet disturbed the pre-existing wave pattern at these longitudes.  

We therefore speculate that the SEBs wave develops when the SEB is generally quiet, lacking in large-scale convective activity (`rifting').  If the wave is generated via the instability mechanisms described above, then the violation of the stability criteria does not necessarily guarantee the development of a growing wave mode, which could explain its ephemeral nature.  Alternatively, there could be some time-dependent interaction between the Great Red Spot and the SEBs jet that has not yet been quantified observationally.  At other times, when the normal rifting west of the GRS creates fluctuating patterns of vorticity and turbulence on the SEBs jet, it may modify conditions so that they are incompatible with the formation of a regular wave pattern.  Furthermore, fluctuations in the temperature structure across the SEBs jet \citep[e.g., those associated with the fade and revival cycle,][]{11fletcher_fade} may alter the amplitude of the $\beta_e$ sign reversal, removing the potential for baroclinic instabilities at certain epochs.  A detailed assessment of thermal changes across Jupiter's belt/zone structure will be the subject of future work, using modern ground-based infrared instrumentation to sound the jovian temperature structure as a function of time.

\section{Conclusions}
\label{conclusion}

Visible light observations have revealed the presence of an ephemeral small-scale wave pattern on Jupiter's fastest retrograding jet at $19.5-20.0^\circ$S, the SEBs.  This wave generally appears when the SEB is relatively quiescent, and is particularly notable when the SEB has undergone a complete fade (whitening).  It appears to be absent when vigorous rifting in the SEB is active.  The wave train is retrograding (moving westward) with a phase speed smaller than the retrograde jet peak, and shows a well-defined correlation between the longitudinal spacing of the wave crests and the phase speed.  When the wave is most closely spaced, it travels with the highest retrograde phase speed.  Extrapolating to zero spacing (i.e., small wavelengths), we find that the wave would be travelling at the speed of the ambient retrograde flow.  The wave amplitude appears to be meridionally confined within $\approx2^\circ$ latitude and vertically confined to approximately a scale height (20 km at 500 mbar), given that near-infrared and thermal-infrared observations sensing $p<300$ mbar failed to detect the wave pattern.

The SEBs differs from Rossby wave activity identified on the jets bounding the North Equatorial Belt.  Specifically, the \textit{eastward} motion of the wave crests with respect to the zonal flow implies that this cannot be a Rossby wave in the classical sense.  Although the retrograde NEBn jet at $17^\circ$N does sometimes host Rossby wave activity, we find that the vertical shear on this jet is much stronger than for the SEBs, which remains retrograde throughout the upper troposphere.  Instead, we use atmospheric temperatures and winds derived during the Cassini encounter to determine the gradient of quasi-geostrophic potential vorticity ($\beta_e$) to trace the dynamics, finding a sign reversal in $\beta_e$ associated with the SEBs jet due to strong vertical and horizontal curvature of the zonal windfield.  This region of negative $\beta_e$ is confined to the 400-700 mbar region, consistent with the vertical confinement of the wave close to the cloud tops.  Such sign reversals are a necessary condition for violation of the Charney-Stern stability criterion, suggesting the possibility for growth of baroclinically unstable modes that could equilibrate to form the observed dispersive wave pattern.  

The discovery of this intriguing new wave pattern has been made possible by the long-term database of observations of amateur observers, allowing the identification of the phase-speed-versus-wavelength relationship.  Further progress in understanding the wave could be provided by a record of the changes in thermal structure (and hence the size of the unstable negative $\beta_e$ region) over time, to understand how the wave pattern responds to thermal changes. These temporally-variable properties of jovian waves can provide new insights into the vertical and horizontal structure of the background atmosphere in which they propagate.

\section*{Acknowledgments}
This study has been made possible by the efforts of numerous amateur observers (listed in reports on the BAA and JUPOS websites) and the JUPOS project developed by Grischa Hahn and Hans-J{\"o}rg Mettig.  Fletcher was supported by a Royal Society Research Fellowship at the University of Leicester.   Orton was supported by NASA from its Planetary Astronomy program through an award issued by the Jet Propulsion Laboratory, California Institute of Technology.  We thank P. Irwin and colleagues for the use of the NEMESIS retrieval algorithm for the original CIRS spectral inversions in 2009, and P. Read for his constructive comments on the dynamic implications of this wave detection.  We also thank two anonymous reviewers for their helpful criticism of this manuscript.  We wish to thank the director and staff of the ESO Very Large Telescope for their assistance with the execution of the VISIR observations. This investigation was partially based on thermal-infrared observations acquired at the Paranal UT3/Melipal Observatory using Directors Discretionary Time (program ID 286.C-5009A) and near-infrared observations acquired at NASAÕs Infrared Telescope Facility (program 2015B010), both led by PI Glenn Orton.  We also express thanks to Padma Yanamandra-Fisher, Gregory Villard and Shirley Trinh for assistance during the IRTF observations, and Michael Sola for initial reduction of the NSFCam2 observations.   The authors wish to recognise and acknowledge the very significant cultural role and reverence that the summit of Mauna Kea has always had within the indigenous Hawaiian community.  We are most fortunate to have the opportunity to conduct observations from this mountain.

\bibliographystyle{elsarticle-harv}
\bibliography{references}







\end{document}